\documentclass[a4paper,11pt]{article}
\pdfoutput=1 

\usepackage{jcappub} 

\usepackage[T1]{fontenc} 

\usepackage{physics}
\usepackage{xcolor}
\usepackage{hyperref}
\usepackage{multirow}
\usepackage{graphicx}
\usepackage{subcaption}
\usepackage{soul}
\usepackage{orcidlink}

\title{Doppler Bias: Impact of Peculiar Velocities on Color Selection and the Large Scale Structure of Galaxy Surveys}

\author[a,b,c,1]{Batia Friedman-Shaw\orcidlink{0009-0008-1840-8481},\note{Corresponding author.}}
\author[c,b,a,*]{Alex Krolewski\orcidlink{0000-0003-2183-7021},}
\author[d]{Matteo Foglieni\orcidlink{0000-0001-8736-4517},}
\author[a,b,c]{and Niayesh Afshordi\orcidlink{0000-0002-9940-7040}}

\affiliation[a]{Perimeter Institute for Theoretical Physics,\\31 Caroline Street North, Waterloo, Ontario, N2L 2Y5, Canada}
\affiliation[b]{Department of Physics and Astronomy, University of Waterloo \\200 University Avenue West, Waterloo, ON, N2L 3G1, Canada}
\affiliation[c]{Waterloo Centre for Astrophysics, University of Waterloo \\
200 University Avenue West, Waterloo, ON N2L 3G1, Canada}
\affiliation[*] {CITA National Fellow}
\affiliation[d]{Leibniz Supercomputing Centre (LRZ), Boltzmannstraße 1, 85748 Garching bei M\"unchen, Germany}

\emailAdd{bfriedmanshaw@perimeterinstitute.ca}
\emailAdd{akrolews@uwaterloo.ca}
\emailAdd{matteo.foglieni@lrz.de}
\emailAdd{nafshordi@pitp.ca}

\abstract{
Lightcone selection effects on cosmic observables must be precisely accounted for in the next generation of surveys, including the Dark Energy Spectroscopic Instrument (DESI) survey. This will allow us to correctly model the data and extract subtle shifts from general-relativistic effects. We examine the effects of peculiar velocities on color selection in spectroscopic galaxy surveys, with a focus on their implications for the galaxy clustering dipole $P_1(k)$. Using DESI Emission Line Galaxy (ELG) targets, we show that peculiar velocities can shift spectral emission features into or out of filter bands, modifying galaxy colors and thereby changing galaxy selection. This phenomenon mimics the effect of evolution bias, and we refer to it as the Doppler bias, $b_D$. The Doppler bias is of comparable size to the evolution bias at $0.8 < z < 1$, where it is largest. This enhances the ELG-LRG (Luminous Red Galaxy) cross-correlation dipole by 25---50\%. This could be detectable at the $\sim$6$\sigma$ level for the full DESI survey. Additionally, we found that our $b_D$ estimate is impacted by the incompleteness of the parent ELG sample. Therefore, this work highlights the essential need for careful consideration of spectral-dependent biases caused by peculiar velocities during the selection phase of galaxy surveys, to enable unbiased analyses.
}

\begin{document}
\maketitle
\flushbottom

\section{Introduction}
We aim to learn about cosmology from the Gaussian linear matter field, but we instead observe
the redshift-space distribution of biased tracers (galaxies), which inhabit the evolved, highly nonlinear density field.
On large scales, most of the cosmological information is contained in the two-point statistics -- the power spectrum or the correlation function -- which are very well understood theoretically on linear and quasilinear scales.
On large scales (i.e.\ not including complications from nonlinearities and higher-order bias), the dominant contributor to the observed
galaxy density, beyond the real-space density itself, are the peculiar velocities through redshift-space distortions (RSD) \cite{Kaiser87}.
There are a number of other sub-leading contributions, generally termed relativistic effects, which have been extensively studied, especially recently as galaxy surveys reach unprecedented precision \cite{DiDioSeljak,DiDioBeutler20,BeutlerDiDio20,Castorina_2022}. These include Doppler effects, magnification bias, gravitational redshift, and the integrated Sachs-Wolfe effect.
These additional effects offer a rich laboratory to test gravity and dark matter interactions on large scales \cite{Saga23,BeltranJimenez23,Umeh21,Bonvin18,Bonvin20}.
Additionally, relativistic effects have a similar scale dependence to the large-scale scale-dependent bias induced by local primordial non-Gaussianities (PNG) \cite{Camera15,Alonso15,Fonseca15,Raccanelli16,Yoo12,Wang20}. Therefore, it is increasingly important to capture relativistic effects in order to better constrain local PNG.
In the redshift-space galaxy density field, the relativistic terms are suppressed relative to the dominant matter density, lensing, and RSD components by at least one power of $\mathcal{H}/k$,
where $\mathcal{H} = aH$ is the comoving Hubble parameter (as described in detail in Section~\ref{sec:doppler_shift_theory} below).
When computing $P(k) = \langle \delta_k \delta_k^* \rangle$, the peculiar velocity terms are therefore
suppressed by a factor of $(\mathcal{H}/k)^2$.
The density-velocity cross-terms are imaginary in Fourier space, and thus cancel when multiplying $\delta_k$ by its conjugate $\delta_k^*$.
However, if we instead consider the cross-correlation of two
tracers with different linear biases $b_A$ and $b_B$,
there is a residual imaginary cross-term $\propto (b_B - b_A) \mathcal{H}/k$ \cite{mcdonald09,Bonvin14}. When computing the standard Legendre multipole moments of the power spectrum, this imaginary term generates a nonzero clustering dipole. Since the standard contributions to this dipole are zero, the multi-tracer dipole offers an ideal opportunity to study relativistic effects.

The amplitude of the Doppler terms (and thus, the clustering dipole) depends on astrophysical bias parameters
as well as the cosmological contributions.
The clustering bias has been extensively studied \cite{Desjacques18}; the large-scale contributions from beyond-linear-bias terms are suppressed (in a standard $\Lambda$CDM cosmology with Gaussian initial conditions) and we will therefore only consider linear bias in this work (though see \cite{Castiblanco19,Castorina_2022,Noorikuhani23} for interactions between higher-order clustering bias and relativistic terms on quasi-linear scales).
The magnification bias describes the response of galaxy number counts to a uniform magnification (e.g.,\ from gravitational lensing or Doppler effects). It can be straightforwardly measured using the faint-end slope of the luminosity function for a simple selection with a one-band flux limit \cite{BartelmanSchneider01}. 
Measuring the magnification bias is more involved for realistic
color-selected galaxy samples \cite{wenzl2023magnification,ElvinPoole23}, but it is nevertheless a straightforward measurement once the galaxy selection function is defined.
Finally, the evolution bias defines the response of the comoving galaxy number density to a peculiar velocity shift, and is nonzero if the underlying number density changes as a function of redshift.
This can arise, e.g.\ from mergers in a sample of halos or from mergers or evolution in the underlying galaxy sample \cite{Maartens22}. In a galaxy sample selected into a redshift bin,
peculiar velocities move galaxies into and out of the sample across the redshift bin boundary. If the galaxy number density changes in redshift, galaxies moving in one direction will be favored, creating a clustering dipole.

Previous derivations of the evolution bias have considered simple situations, e.g.\ a single-band flux cut or line lumnosity limit in \cite{Maartens22}. However, a realistic galaxy selection function leads to a more complicated impact of peculiar velocities on the galaxy sample. We consider a spectroscopic survey, where galaxies are targeted based on their fluxes in multiple broad bands, then kept in the sample if spectroscopically confirmed to lie in the correct redshift range.
Thus, in addition to perturbing the observed redshift away from the true redshift, peculiar velocities also modify the broad-band fluxes by shifting the galaxy spectrum. This can also affect galaxy number counts by shifting galaxies into, or out of, the initial color selection box. Although this may be a small effect, it can be enhanced if the galaxy flux is dominated by one or a few narrow emission lines, e.g.\ the case with Emission Line Galaxies (ELGs).

We study this effect of peculiar velocities impacting the color selection of galaxies, which we call the Doppler bias, in detail
for a particular set of Emission Line Galaxies following the DESI
\cite{desicollaboration2016desi} ELG target selection \cite{Raichoor23}.
We build on the work of \cite{Maartens22}, replacing the assumption of a global, redshift-dependent K-correction
with the full diversity of observed galaxy spectra; and using a multi-band galaxy selection rather than a single flux cut.\footnote{A similar method was applied to estimate the impact of peculiar velocities on the clustering dipole of BOSS LRGs in \cite{Alam18a}. This paper directly measured the number of galaxies which scatter across the flux boundary using peculiar velocities estimated by reconstruction, without ever quoting a value for the evolution bias. On the other hand, we measure the change in evolution bias to better connect to the theory in \cite{DiDioSeljak,DiDioBeutler20,Beutler_2019,Castorina_2022}.}
We show that, for the DESI ELG sample, the Doppler bias is similar in magnitude to the evolution bias from the change in comoving number density, and thus needs to be considered to accurately 
model relativistic effects.
This effect will be increasingly important for future work, since ELG samples are increasingly favored for high-redshift galaxy surveys (H$\alpha$ for Euclid \cite{laureijs2011euclid} and Roman \cite{wfirst}; Lyman Alpha Emitters for DESI-II \cite{Schlegel22a} and MegaMapper \cite{Schlegel22b}).

We organize this paper as follows. Section \ref{Theory} is an overview of the theory describing the impact of relativistic effects on galaxy number counts, and provides an overview of our synthetic-sample approach to measuring the Doppler bias. Section \ref{Selection details} describes the details of the DESI data and galaxy selection. Section \ref{doppler bias} details our measurement of the Doppler bias,
while
Section \ref{bias measuring}
presents measurements of the magnification bias and evolution bias. Finally, we show the impact on galaxy clustering in Section~\ref{two-point stats}. Lastly, we provide concluding remarks in Section \ref{conclusion}.

Unless noted otherwise, we set the speed of light $c=1$ and use $h^{-1}$Mpc as units of both space and time.

\section{Theory}
\label{Theory}

\subsection{Impact of the Doppler shift on the observed galaxy power spectrum}
\label{sec:doppler_shift_theory}
In linear theory (including relativistic corrections), the galaxy number counts $\Delta_g$\footnote{Note that we are intentionally using the observed galaxy number counts, $\Delta_g$, as opposed to the true galaxy overdensity field, $\delta_g$, following the prevailing literature on relativistic effects. See \cite{Bonvin_2011} for more information.}, selected only by a {\it bolometric} flux cut, can be related to the matter overdensity $\delta_m$ and the line-of-sight velocity $v_{\parallel}$
\citep{DiDioSeljak,Foglieni23} 
\begin{align}
    \Delta_g &= \bigg\{b_1 \delta_m + \mathcal{H}^{-1}\partial_r v_{\parallel}\bigg\} + \left\{\frac{5s-2}{2}\int_0^\chi d\chi' \frac{\chi - \chi'}{\chi \chi'} \Delta_\Omega (\Psi + \Phi)
    \right\} \nonumber\\&+ \bigg\{ \mathcal{R}_v (v_{\parallel} - v_{\parallel, o}) - (2 - 5 s) v_{\parallel, o} \bigg\} 
     \nonumber\\&+ \left\{ \left( \mathcal{R}_v - \frac{2-5s}{\mathcal{H}_0 \chi}\right) \mathcal{H}_0 V_o + (\mathcal{R}_v + 1) \Psi - \mathcal{R}_v \Psi_o + (5 s -2) \Phi 
    + \dot{\Phi} \mathcal{H}^{-1} + (b_{e} - 3) \mathcal{H} V \right\}
    \nonumber\\& + \left\{ \frac{2-5s}{\chi} \int_0^\chi (\Psi + \Phi) d\chi' + \mathcal{R}_v \int_0^\chi (\dot{\Psi} + \dot{\Phi}) d\chi'\right\}
\label{eqn:delta_g_relativistic}
\end{align}
where 
\begin{equation}
    \mathcal{R}_v = 5s + \frac{2 - 5s}{\mathcal{H}\chi} + \frac{\dot{\mathcal{H}}}{\mathcal{H}^2} - b_{e}.
    \label{eqn:R}
\end{equation}
In Eq. \ref{eqn:delta_g_relativistic},
$b_1$ is the linear bias;
$s$ is the magnification bias;
$\Psi$ and $\Phi$ are the time-time and space-space parts of the perturbed gravitational potential;
the subscript ``$o$'' indicates the peculiar velocity
and potentials at the position of the Earth (observer); $V$ is the velocity potential defined via $\vec{v} = - \vec{\nabla}{V}$; and overdots are partial time derivatives. $\mathcal{R}_v$ follows the notation in \cite{Foglieni23}, but a subscript $v$ has been added for clarity to distinguish $\mathcal{R}_v$ from the comoving curvature perturbation.

The terms in Eq. \ref{eqn:delta_g_relativistic} are split the same way as \cite{Foglieni23}: Newtonian term (linear bias plus linear RSD); lensing; Doppler; local gravitational potentials;
and integrated gravitational potentials. Each of these six terms appears respectively in the equation with six sets of brackets.

The biases $b_1$, $s$, and $b_{e}$ depend on the galaxy
selection function. $b_1$ is the linear Eulerian galaxy clustering bias; $s$ is the magnification bias, defined as the response of the galaxy comoving number density, $\bar{n}$, to an infinitesimal change in the survey limiting magnitude $m$
\begin{equation}
    s \equiv \frac{\partial \log_{10}{\bar{n}}}{\partial m}
    \label{eqn:mag_bias}.
\end{equation}
Recalling the conversion from magnitude to luminosity, this is identical
to Equation 2.10 in \cite{DiDioSeljak}.
Next, $b_{e}$ is the evolution bias, the response of $\bar{n}$ to an infinitesimal perturbation in redshift\footnote{Some authors define $f_\textrm{evo}$ as our $b_{e}$ plus 3; the extra factor of 3 accounts for the difference between comoving and physical number density. When using a comoving number density, $b_{e} = 0$ for dark matter, and when using a physical number density, so that $\rho \propto a^{-3}$, $b_{e} = 3$ \citep{BeutlerDiDio20}.}
\begin{equation}
    b_{e} = - (1 + z) \frac{d \ln{\bar{n}}}{dz}.
    \label{eqn:ev_bias}
\end{equation}
As emphasized in \cite{Maartens22}, $b_e$ and $s$ are defined using partial derivatives that, respectively, hold the flux cut and the redshift constant. We follow this approach in our calculation of the Doppler bias in the next section, only varying the infinitesimal peculiar velocity and otherwise holding the selection criterion constant.

An important assumption of Eqs.~\ref{eqn:mag_bias} and~\ref{eqn:ev_bias} is that the response of the galaxy number density to redshifting and magnification is linear. This must be true for a sufficiently small change in magnitude or redshift, and has been explicitly tested for measurements of magnification bias in \cite{wenzl2023magnification}. For the Doppler bias, we perform a similar test below, exploring the smallness of the stepsize. We furthermore note that the number density response should be linear on the characteristic scales of linear peculiar velocities, $v_{\textrm{rms}} \sim$ 150 km s$^{-1}$. If this is not satisfied, there could be higher-order corrections in Eq.~\ref{eqn:delta_g_relativistic}, modifying the relationship between the Doppler bias and the observable galaxy power spectra.

\begin{figure}[htbp] 
\centering
\includegraphics[width=\textwidth]{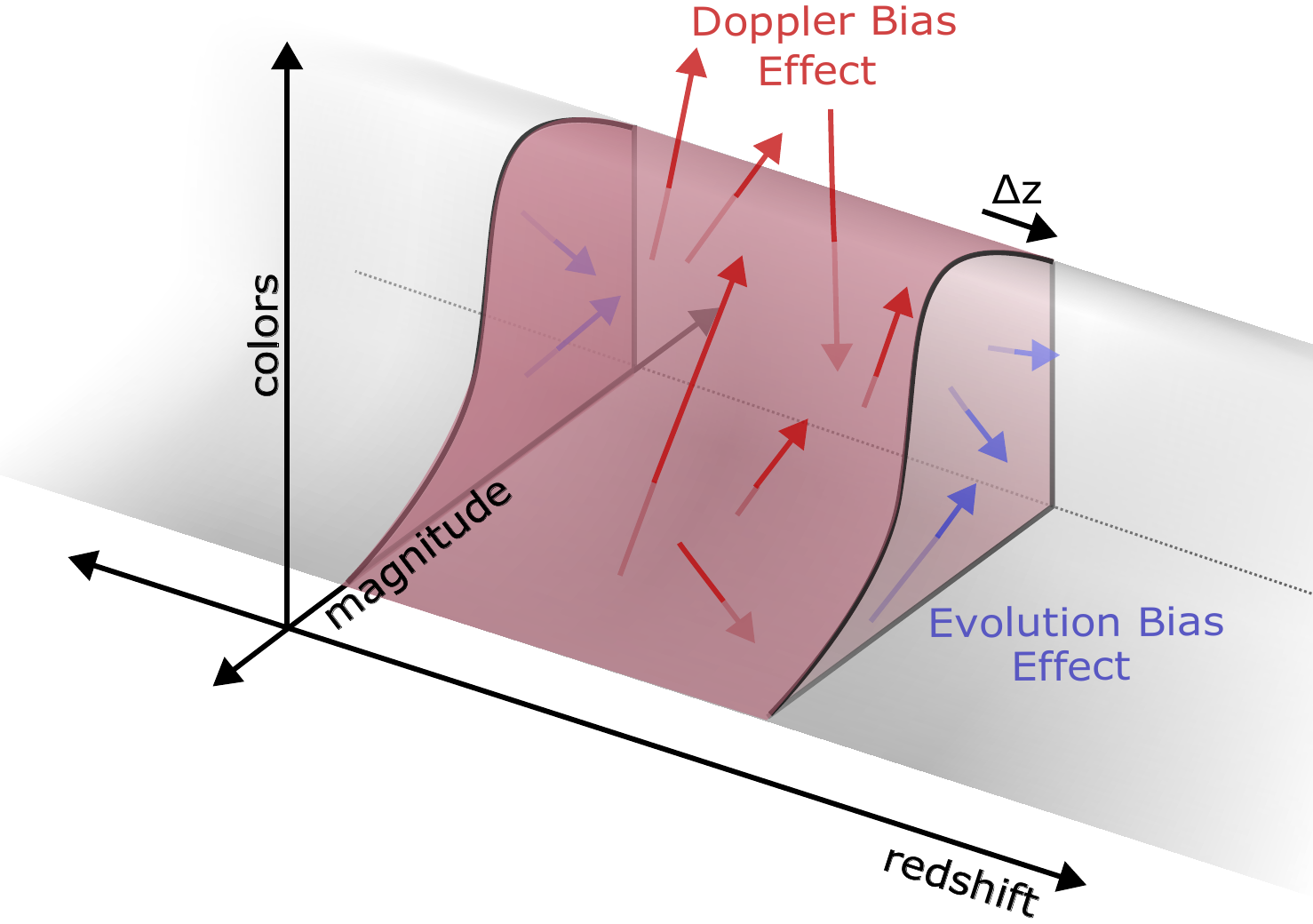}
    \caption{Schematic description of the Doppler bias computed in this paper, compared to the evolution bias defined in previous work. In this plot, the galaxies are selected inside the pink shaded region -- they fall within color and magnitude bounds, as well as inside a redshift bin. 
    The two axes of colors and magnitude 
    stand in for the seven-dimensional space of colors and magnitude from the $g$, $r$, $z$, and $g_{\textrm{fiber}}$ magnitudes. Slices through this space are shown by the selection box on the left hand side of Fig.~\ref{fig:bd_features}. A fixed line-of-sight peculiar velocity, $v_r$, leads to a redshift change $\Delta z= (1+z)v_r/c$ as shown by the arrow above the plot. This $\Delta z$ causes an evolution bias which shifts galaxies in and out of the redshift bin, and is represented by the purple arrows that only affect objects at the edges of the redshift boundary. The Doppler bias (represented by red arrows), on the other hand, shifts galaxies in and out of the surface of color and magnitude cuts, and therefore changes the galaxy sample in a distinct manner from the evolution bias. In practice, we compute the change in galaxy number, $\Delta N$, from the evolution bias and the Doppler bias separately, and add them together.}
    \label{fig:b_e_vs_bd}
\end{figure}

\begin{figure}[htbp] 
\centering
\includegraphics[width=1.0\textwidth]{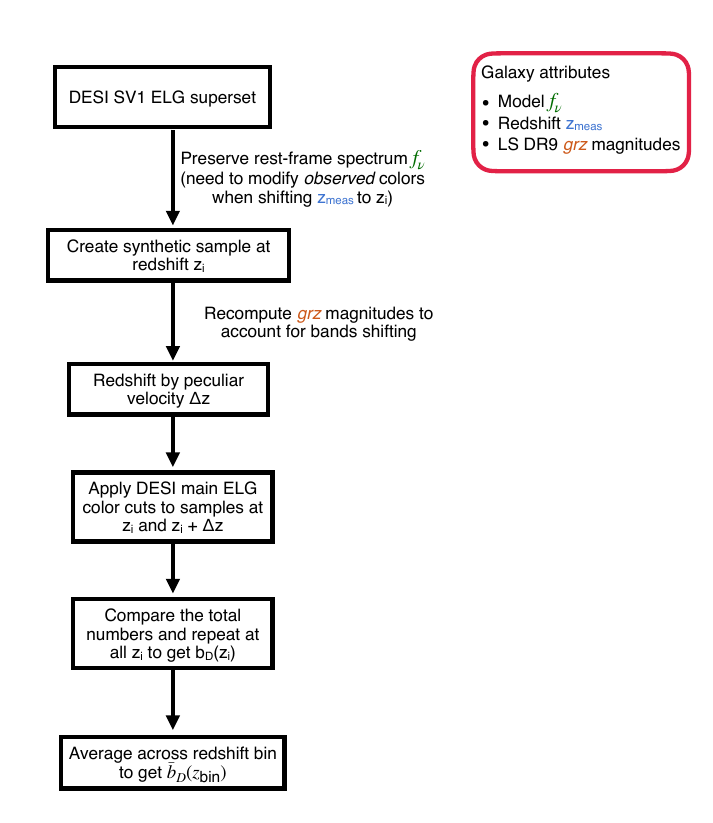}
    \caption{Flowchart summarizing our methodology. We start with the DESI SV1 ELG superset and use it to create synthetic samples at many closely spaced $z_i$, using real galaxies in the range ($z_i -0.05$, $z_i$) and preserving their rest-frame flux when shifting them to $z_i$. We then apply a small peculiar velocity to this sample, recompute its $grz$ magnitudes, and compare the number of DESI main ELGs selected from the superset with and without the peculiar velocity. We then average within redshift bins to compute $\bar{b}_D(z_{\rm bin})$, which can be directly related to LSS observables.}
    \label{fig:flowchart}
\end{figure}

The peculiar velocity biases described above (i.e.\ evolution bias and other terms in Eq.~\ref{eqn:R})
break isotropy in the galaxy field, and lead to a nonzero dipole in the power spectrum and correlation function as their effects only impact the line of sight direction.
Another major contributor to the dipole (and other odd multipoles) comes from wide-angle effects due to the galaxy power spectrum estimator definition with respect to the line of sight.
Within the Yamamoto estimator for the galaxy power spectrum multipoles
\cite{Yamamoto}, the line-of-sight direction $\hat{\mathbf{d}}$ can be chosen separately for each galaxy pair with position vectors $\mathbf{s}_1$ and $\mathbf{s}_2$:
\begin{equation}
   \hat{P}_{\ell} (k) = \left\langle \frac{2 \ell + 1}{2A} \int \, d\mathbf{s}_1  \int \, d\mathbf{s}_2  \,\delta_g(\mathbf{s}_1) \delta_g(\mathbf{s}_2) e^{i \mathbf{k} \cdot (\mathbf{s}_1 - \mathbf{s}_2)} \mathcal{L}_{\ell}(\hat{\mathbf{k}} \cdot \hat{\mathbf{d}}) - S_\ell \right\rangle
   \label{eqn:yamamoto}
\end{equation}
where $\mathcal{L}_\ell$ is the Legendre polynomial of order $\ell$, $A$ is the normalization and $S_\ell$ is the shot noise.
If the line-of-sight is chosen in a symmetry-breaking way, so that it is the same as the unit vector pointing to one of the galaxies, then the power spectrum can be calculated using fast Fourier transform methods which scale
favorably with the large number of galaxy pairs in modern galaxy surveys
\cite{Bianchi15,Scoccimarro15,Slepian15,Hand18}.
This is referred to as the end-point line of sight definition in \cite{Beutler_2019}, in contrast to more symmetrical line-of-sight definitions
that use the angular bisector or mean galaxy position vector.

However, breaking the symmetry between galaxies in a pair breaks parity symmetry and thus leads to a nonzero dipole \cite{BeutlerMcDonald21}.
The linear RSD term arising in the galaxy power spectrum assumes
that the angle separating the galaxies is small. On large scales, this assumption breaks down
\cite{Szalay98,Szapudi04,Papai08,Raccanelli10,Yoo14,Castorina18a,Castorina18b}, and the wide-angle contributions are especially large for the end-point line of sight choice \cite{Reimberg16,Castorina18a,Castorina18b,Beutler_2019}. This contribution to the dipole and the octopole can be as large or larger than the 
contribution from peculiar velocities.

Derivations of the magnification and evolution biases often assume
a simple single-band flux or luminosity cut to define the galaxy sample.
However, realistic galaxy samples, particularly in spectroscopic
surveys, are much more complicated, with multiple flux and color cuts, often in different apertures (to eliminate failed redshifts of low surface-brightness galaxies or remove stars with point-source morphology).
The impact of these subtleties on magnification bias is studied in \cite{wenzl2023magnification}. Our paper extends similar considerations to the evolution bias.

\subsection{Synthetic sample approach to measuring the Doppler bias}
As with the magnification bias, the evolution bias
must be defined for a specific multi-color galaxy selection, using the underlying galaxy spectra to determine the effects of peculiar velocity on the galaxy selection function.
This means that evolution bias is more complex than the typical description of ``derivative of the number density with respect to redshift.''
Peculiar velocities affect galaxy fluxes as well as redshifts,
by shifting emission features in and out of the bandpass
used to define the flux. We therefore split the evolution bias into two parts: the previously-defined evolution bias $b_e$
defined as the redshift derivative of the true comoving number density;
and the impact of the Doppler shift on the observed fluxes, which we call the Doppler bias, $b_D$. Fig.~\ref{fig:b_e_vs_bd} shows this schematically: evolution bias shifts the galaxy only right or left (changing the redshift) whereas Doppler bias can also shift the galaxies up and down (changing the fluxes). Unlike magnification bias, Doppler bias changes relative fluxes and thus colors, leading to additional effects from the color selection.

Similar effects have been considered previously, for instance in Refs.~\cite{Maartens22,Alam18a}.
Our work differs in a few respects.
Ref.~\cite{Maartens22} considers the impact of K-corrections on the evolution bias for a single-band-selected galaxy survey, due to the fact that the peculiar velocity shifts
a different part of the spectrum into the observed band.
We go beyond this treatment in two respects: first, they calculate K-corrections assuming all galaxies have
the same spectrum, and second, we consider a realistic galaxy color selection, which is far more complicated than a simple flux cut.
Ref.~\cite{Alam18a} considers a very similar change in the CMASS galaxy selection induced by peculiar velocities; however,
they directly estimate which galaxies scatter into and out of the sample based on peculiar velocity reconstruction,
rather than expressing the effect as an evolution-like bias like we do. Moreover, they consider the CMASS LRG sample, whereas we consider the DESI ELG sample which is much more susceptible to the Doppler bias.

Here we define the Doppler bias similarly to the evolution bias except $N_{\textrm{color}}$ now refers to the number of galaxies within the color and magnitude selection cut:
\begin{equation}
    b_{D} \equiv -\qty(1+z)\frac{\ln N_{\textrm{color}}(z + \Delta z_\textrm{pec}) - \ln N_{\textrm{color}}(z)}{\Delta z_\textrm{pec}}.
    \label{eq:b_doppler}
\end{equation}
Note that in this equation, we use the total number of galaxies within an observed redshift bin, $N_{\textrm{color}}(z)$, rather than the number density (number per comoving volume). The derivative is taken with respect to the line of sight peculiar velocity, $v_\parallel = \Delta z_{\rm pec}/(1+z)$,  rather than redshift. Therefore, in $b_D$, the redshift dependence of comoving volume does not contribute to the derivative, whereas in $b_e$, where we take a redshift derivative, it does contribute, and we must use the comoving number density $\bar{n}$.

In Fig.~\ref{fig:flowchart}, we present a flowchart summarizing our methodology. We begin with a superset sample, described in Section~\ref{Selection details},
intended to contain all possible galaxies that could be part of the galaxy selection under consideration (the DESI ELG main survey selection). 
For each galaxy, we have a synthetic spectrum (the best-fit model spectrum as determined by the DESI redshift-fitting software), and integrate that spectrum to measure fluxes to determine how peculiar velocities change the galaxy colors.
We then create synthetic samples at different redshifts ($z_i$), preserving the rest-frame flux for each galaxy, but moving galaxies to a common redshift to increase the sample size at each redshift and prevent confusion with redshift-dependent selection effects in the underlying superset. We then redshift all the galaxies under consideration by a small peculiar velocity $\Delta z$, compute the change in galaxy number density, and finally sum the change in number density over a specified bin in observed redshift to measure the Doppler bias $b_D$.

When including this term in our power spectrum calculations, we add $b_D$ to all instances of $b_e$ as they factor into the physics in the same way. Therefore $\mathcal{R}_v$ in Eq. \ref{eqn:R} becomes 

\begin{equation}
    \mathcal{R}_v = 5s + \frac{2 - 5s}{\mathcal{H}\chi} + \frac{\dot{\mathcal{H}}}{\mathcal{H}^2} - b_{e} - b_{D}.
    \label{eqn:b_D_correction}
\end{equation}

\subsection{Power Spectrum Dipole Calculation}

The theoretical power spectrum multipoles computed in this work were calculated using GaPSE,\footnote{\url{https://github.com/foglienimatteo/GaPSE.jl}} \cite{Castorina_2022,Foglieni23,Pantiri24} a Galaxy Power Spectrum Estimator code written in Julia. More specifically, GaPSE computes all the full relativistic galaxy auto and cross correlation functions for an arbitrary multipole order, using the endpoint line of sight employed in the Yamamoto estimator. The correlation functions can be computed with or without an arbitrary window function. 
To speed up the evaluations, GaPSE takes advantage of an effective redshift approximation, which is shown to be accurate within $\sim$1\% (Fig.~4 of \cite{Castorina_2022}) for narrow bins like the ones we considered.

\begin{table}[]
    \centering
    \begin{tabular}{ |p{3cm}||p{3cm}|p{3cm}|  }
     \hline
     ELG Biases & North & South\\
     \hline
     $b_1$   & 1.3    & 1.3\\
     $b_e$&   $-1.2$  & $-2.1$\\
     $s$& 0.44 & 0.38\\
    $b_D$& $-4.7$ & $-5.9$\\
     \hline
      LRG Biases & North & South\\
     \hline
     $b_2$& 2  & 2\\
     $b_e$& 8.1  & 10.2\\
     $s$& 1 & 1\\
     $b_D$& 0 & 0\\
     \hline
    \end{tabular}
    \caption{Biases used in the computations made with GaPSE for Emission Line Galaxies and Luminous Red Galaxies. These are averaged (weighted by number density) bias values over a redshift bin from 0.8 to 1.0. Note that different values were used for the North and South samples as each sample is slightly different.
    }
    \label{tab:all_biases}
\end{table}

We match the fiducial cosmology of the DESI collaboration \cite{DESI23a}, from the 
Planck 2018 TT,TE,EE+lowE+lensing marginalized means with a single massive neutrino of mass 0.06 eV \cite{planck2018}:
$\Omega_b h^2 = 0.02237$, $\Omega_c h^2 = 0.1200$,
$\ln 10^{10} A_s = 3.044$, $n_s = 0.9649$, $H_0 =  67.36$ km s$^{-1}$ Mpc$^{-1}$.

\begin{figure}
    \centering
    \includegraphics[width=\textwidth]{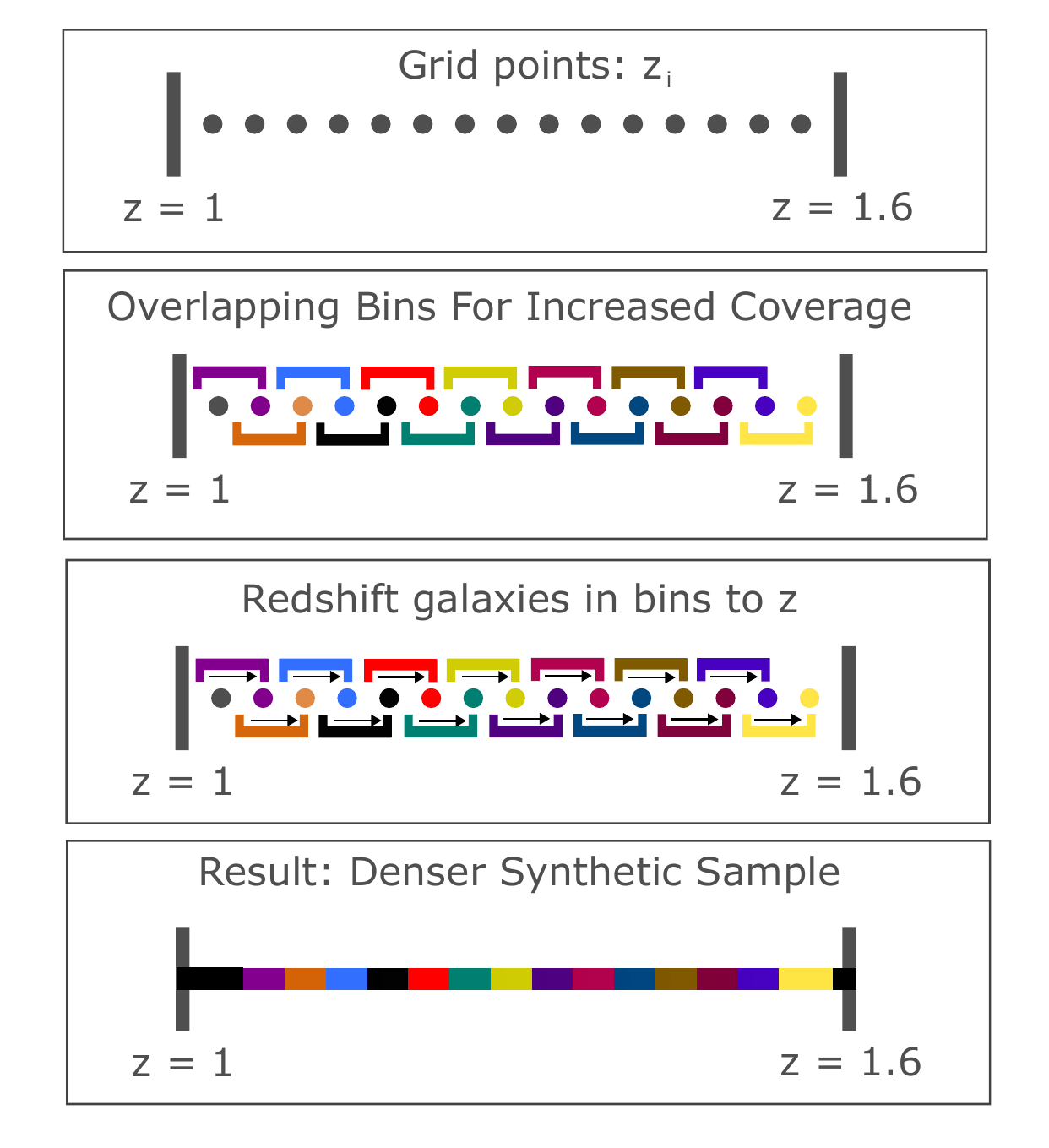}
    \caption{\textbf{Schematic illustrating the synthetic sample creation, for which we used a default bin size of 0.05.}}
    \label{fig:synth_sample}
\end{figure}

\section{DESI Emission Line Galaxy selection}
\label{Selection details}
\begin{figure}[htbp]
    \centering
    \begin{subfigure}[b]{0.45\textwidth}
        \includegraphics[width = \textwidth]{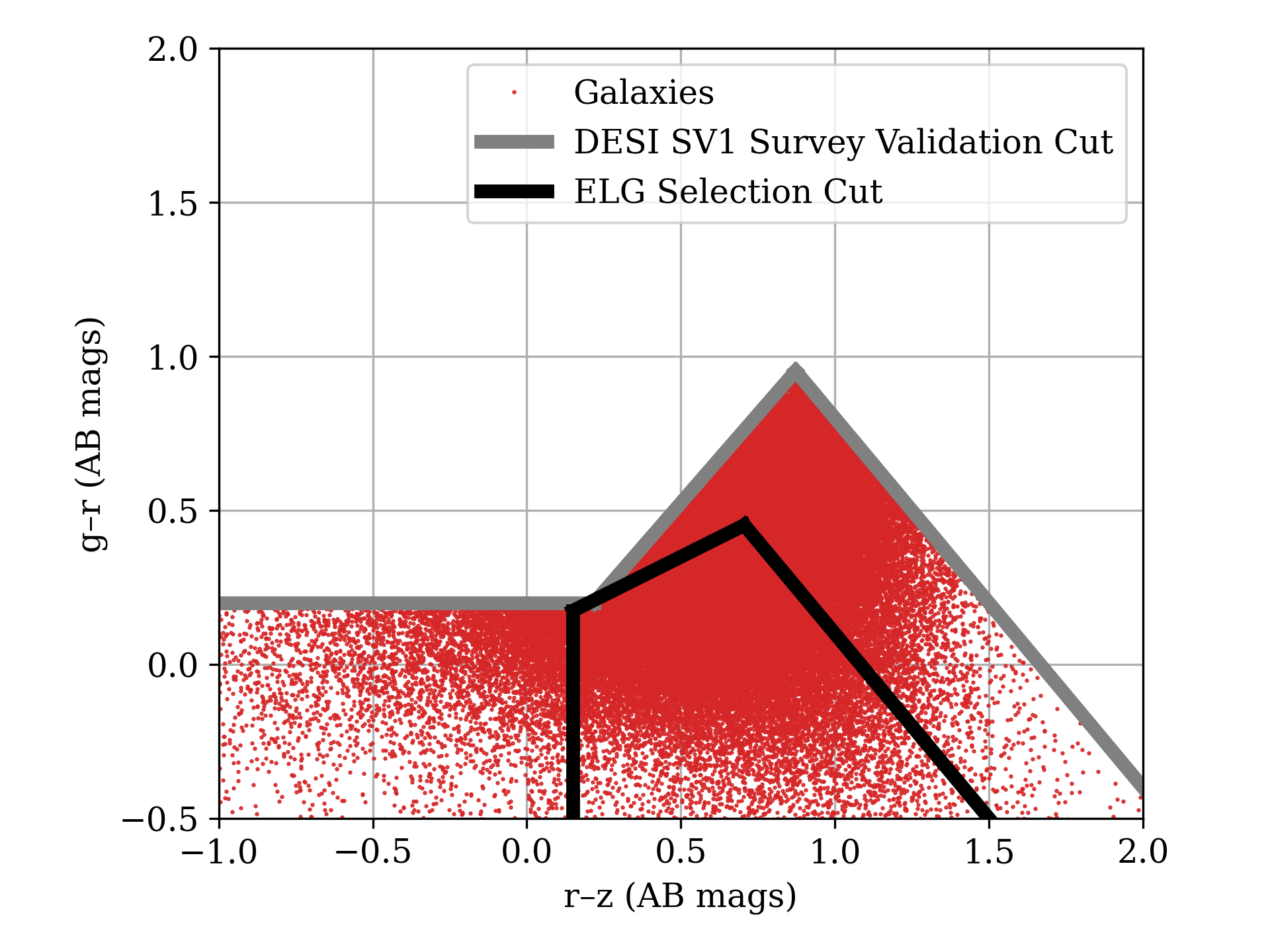}
    \end{subfigure}
    \begin{subfigure}[b]{0.45\textwidth}
        \includegraphics[width=\textwidth]{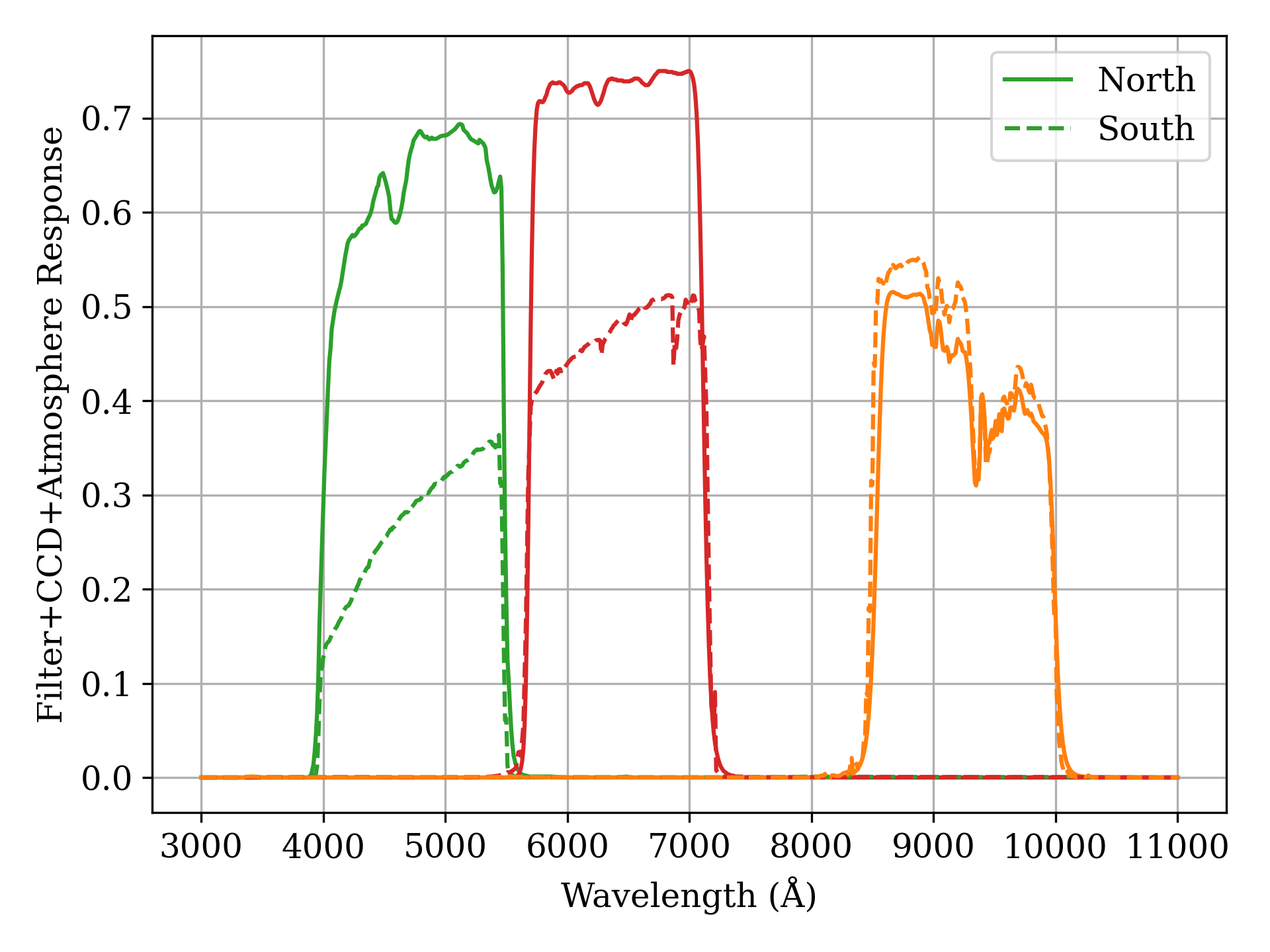}
    \end{subfigure}
    \caption{\textit{Left:} distribution of DESI ELGs in $grz$ space. More detail in Fig. 3 of \citep{Raichoor23}. The grey line shows the superset defined by 
    early Survey Validation (SV1), whereas the black line shows the main survey ELG selection cut. We consider the SV1 ELGs as a superset within which galaxies can scatter into and out of the main survey ELG selection.
    Extinction correction has been applied to the galaxy magnitudes. \textit{Right:} filter response curves for $grz$ bands in the different imaging regions, BASS and MzLS (North; solid) and DECaLS (South; dashed). Filter curves include the entire system throughput (camera, atmosphere, telescope).}
    \label{fig:filters}
\end{figure}

We measure the Doppler and evolution biases using the DESI Emission Line Galaxy (ELG) sample released with the publicly available DESI
Early Data Release (EDR) \citep{DESI23b}\footnote{\url{https://data.desi.lbl.gov/doc/releases/edr/}}. DESI is a five-year survey
targeting 40 million ELGs, Luminous Red Galaxies (LRGs), quasars,
and low-redshift bright galaxies (BGS)
at $0 < z < 2$ \citep{DESI16b,DESI_Instrument,Schlafly23}.
DESI measures 5000 redshifts at once using robotic fiber positioners across a 7.5 deg$^2$ field-of-view, with spectral
resolution and wavelength coverage designed to measure
redshifts from the [OII] doublet at $0.6 < z < 1.6$ \citep{DESI16b,Silber23,Miller23}.
The spectra are processed and reduced \citep{Guy23}, and
redshift fitting is performed automatically using the \texttt{Redrock} software \citep{Bailey23}.

The DESI EDR includes data taken for the Survey Validation (SV) phase \citep{DESI23a} before the main five-year survey,
in order to finalize targeting cuts, test observing strategy, and measure the linear galaxy bias for each sample.
The first part of SV (SV1) was designed to finalize DESI targeting \citep{Myers23}: it uses looser target selection and longer integration times to achieve a high success rate on a superset
of the final DESI targets.
We use the superset defined by SV1 to measure
the magnification, evolution, and Doppler biases for the main ELG selection. Measuring these biases requires
a complete sample to take the derivative  of the selection function
with respect to redshift or luminosity.

We also use the One-Percent Survey of the third portion of SV (SV3) to measure galaxy clustering (and hence infer the linear bias) using data with very high fiber-assignment completeness.
After DESI targets are selected, they must be assigned fibers \citep{Raichoor23}; due to the limited reach of each robotic fiber positioner, fiber-assigned targets cannot be closer than the 1.5 arcmin fiber patrol radius. This leads to distorted small-scale clustering
as well as an incomplete sample, particularly for the ELGs,
which have a target density of 2400 deg$^{-2}$ and are expected to be only $\sim$80\% complete even after the five-year, five-pass main survey.
A variety of methods have been proposed to eliminate the effects of fiber collisions from the main survey data
\citep{Bianchi18,BianchiPercival,Hahn17,Smith19,Pinol17,Pinon24}.
In SV3, 20 separate fields are observed in a rosette pattern with 12--13 overlapping tiles (compared to the 5 overlaps in the DESI main survey), leading to $\sim$95\% fiber-assignment completeness for the ELG sample \citep{DESI23b}.
Using the SV3 data allows us to infer the linear bias directly from the clustering measurements without complicated corrections for incompleteness and fiber-assignment.

The DESI ELG sample is selected from the 3-band $grz$ optical imaging of DESI Legacy Surveys Data Release 9 \citep{Schlegel2021DESI}.
Data Release 9 consists of several observing programs: data north of declination $32.375^{\circ}$
comes from the Beijing-Arizona Sky Survey (BASS) \citep{Zou_2019}
for $g$ and $r$-bands, and the Mayall $z$-band Legacy Survey (MzLS) for $z$-band.
The rest of the imaging comes from DECam \citep{Flaugher_2015} on the 4-m Blanco telescope at CTIO, either using
Dark Energy Survey (DES) data in its 5000 deg$^2$ footprint, or using the DECam Legacy Survey (DECaLS) where DES is not available \citep{Dey_2019}. 
When computing the impact of redshifting on the broad-band colors, we convolve the measured DESI spectra with the appropriate filter curves for BASS, MzLS, and DeCALS,\footnote{Available at \url{https://www.legacysurvey.org/dr9/description/}} as shown in Fig. \ref{fig:filters}.
Data reduction and photometry is performed using the \verb|legacypipe|\footnote{\url{https://github.com/legacysurvey/legacypipe}} and \verb|Tractor| software \citep{dlang2016},
and photometry is measured using AB magnitudes.
Each source is modelled with a point-source, exponential,
de Vaucouleurs or S\`ersic profile, and both total fluxes and fiber fluxes (flux within a 1.5 arcsec DESI fiber convolved with 1 arcsec Gaussian seeing) are reported \citep{Raichoor23}.
The depth varies between footprints: in the North, the 5$\sigma$ $g$-band depth is $g=24.1$,\footnote{Measured for a 0.45 arcsec exponential profile, as reported in Table 1 of \citep{Raichoor23}.} just above the limiting magnitude of ELG selection of $g_{\textrm{fib}} = 24.1$, whereas the depth is $g=24.5$ in the South-DECaLS region and $g=24.9$ in the South-DES region. Here the subscript ``fib'' refers to the $g$ magnitude measured within a 1'' DESI fiber, as opposed to the total magnitude, which is given without a subscript.
We always use extinction-corrected photometry, following DESI target selection and using the extinction map of \cite{SFD98}. While the impact of updated extinction corrections on the ELG target selection may be substantial \cite{Zhou24},
we choose to continue to use the SFD extinction map for consistency with the original DESI target selection.

DESI ELGs are selected using a simple color cut on $g-r$ and $r-z$ colors, and a cut on $g_{\textrm{fib}}$. 
We study the \texttt{ELG\_LOP} selection, as shown in Table 2 of \cite{Raichoor23}, which is designed to target ELGs at $1.1 < z < 1.6$, though in practice selects ELGs over a broder redshift range, $0.6 < z < 1.6$. The ELG selection that we use is given by the following color cuts:
\begin{align}
\begin{split}
r-z &> 0.15 \\
g -r &< 0.5 \times (r - z) + 0.1 \\
g -r &< -1.2 \times (r - z) + 1.3 \\
g &> 20 \\
g_{\textrm{fib}} &< 24.1
\end{split}
\label{eqn:elg_selection}
\end{align}

The SV1 selection extends the color selection in all directions, as shown in Figure 4 of \cite{Raichoor23} and
listed in Table A.1 of \cite{Raichoor23}. 
The color cuts
are the same in the North and South imaging regions, even
though the photometric systems are slightly different.
For clarity, we reproduce the distribution of the DESI ELGs in $grz$ color space in Fig.~\ref{fig:filters}, and also show the SV1 superset and the main survey ELG selections.

We start by selecting all 181,144 SV1 ELGs.
We then remove ELGs with unreliable spectroscopic
redshifts, as measured by a simple cut using the [OII] flux signal-to-noise FOII\_SNR and the $\Delta\chi^2$ between
the best-fit redshift and the second-best fit \citep{Raichoor23}:
\begin{equation}
    \log_{10}(\textsc{FOII\_SNR}) + 0.2 \log_{10}(\textsc{DELTACHI2}) > 0.9
\end{equation}
This leaves 103,249 ELGs.
We also remove any spectra classified as quasars by the quasar afterburner pipeline \citep{Chaussidon_2023},
leaving 98,318 ELGs.
This number is similar to (but slightly different) from the 
number of SV1 ELGs reported in Table 6 of \cite{DESI23b}, due
to the different criterion used for a good redshift.
40,098 ELGs are located in the North, and 58,220 are in the South. Throughout this paper, we compute $b_D$ separately in the North and the South. In principle, since the imaging in DES is deeper, this could lead to a different value of the Doppler bias in the South between DES vs.\ non-DES, since $b_D$ is sensitive to the color distribution of the galaxies. However, the DESI SV3 coverage of DES is sparse, with only 5 out of 80 SV1 tiles in DES (Fig.~1 in \cite{DESI23a}). Within this limited comparison, we find consistency in the Doppler bias values between South and DES, but with large errors on the Doppler bias from DES due to the small sample size.

\texttt{Redrock} returns both the measured spectrum and a best-fit model spectrum; we use the best-fit model spectrum to compute synthetic magnitudes and thus the Doppler bias.
We additionally remove
814 galaxies where the model fluxes are negative or missing. 99 of these galaxies are actually stars and the galaxy template for stars is missing. The other 715 are cases where the template flux is negative, which is allowed by \texttt{Redrock} to model low signal-to-noise spectra; in some of these cases, the observed spectra can be negative due to sky-subtraction issues. We tested recovering these 715 galaxies by adding a constant offset to enforce that the template is always positive. Since we only care about the difference between the non-redshifted and peculiar-velocity redshifted magnitudes, the constant offset should make a minimal impact on the difference.
We find no difference in Fig.~\ref{fig:bd} when we make this change.
Since it is an ad hoc correction, by default we remove these 715 galaxies from our sample.

\section{Measuring the Doppler bias}
\label{doppler bias}

Our process for measuring the Doppler bias involves a number of steps, and is summarized in the flowchart in Fig.~\ref{fig:flowchart}. The basic philosophy is to construct a synthetic sample of ELGs at each redshift ($z_i$) across a finely spaced grid between $z_0 = 0.6$ and $z_n = 1.6$, with sufficient sample size to accurately measure the Doppler bias, and then apply the DESI ELG selection cuts to the synthetic sample.

\subsection{Creating the synthetic sample at each redshift}

The creation of our ELG synthetic sample has the following steps (as illustrated in Fig.~\ref{fig:synth_sample}):

\begin{enumerate}
    \item We start with a measured ELG sample.
    \item We define a fine redshift grid $z_i$.
    \item For each $z_i$, we take galaxies in a redshift range $z_i - 0.05$ to $z_i$.
    \item We redshift their model spectra to $z_i$.
    \item We use this synthetic sample to compute the Doppler bias.
\end{enumerate}

Our ELG synthetic samples consist of galaxies shifted to a grid of redshift values $z_i$. On this grid, the bins of galaxies that are shifted to these $z_i$ overlap significantly such that nearby galaxies are in some sense copied to nearby redshift locations to cover our overall redshift range from $z=0.6$ to $z=1.6$ more completely. When constructing our ELG synthetic samples, we start by redshifting all galaxies within a particular redshift range to $z_i$. We refer to the original redshift of each galaxy as $z^{\textrm{meas},j}$ and each redshift under consideration on the finely spaced grid as $z_i$.\footnote{We neglect redshift uncertainties in this procedure; this is self-consistent because we always use the synthetic spectrum to determine magnitudes. The only redshift error which could affect this procedure is a catastrophic redshift error, which are $<0.2\%$ of the sample \cite{Raichoor23}.}
The superscript $j$ refers to the galaxy under consideration, while $i$ refers to the particular redshift in our well populated superset.

We shift to $z_i$ for two major reasons. First, we want to measure the Doppler bias in very narrow bins of redshift $z_i$, since the [OII] emission line is narrow and the impact of redshifting it may change rapidly across a small change in redshift.
However, this would lead to very narrow bins of ELGs, with very few galaxies in each bin, making it difficult to measure the Doppler shift well. 
Second, using galaxies only at their measured redshifts could introduce observational biases from instrument sensitivity and atmospheric interference. By shifting galaxies to a uniform grid, we create a sample where redshift-dependent effects are minimized.
It may be that the SV1 superset observes galaxies less efficiently at certain redshifts (e.g.,\ due to variations in the DESI instrument  or atmospheric lines \cite{Yu24}), and could have peculiar velocity selection effects already imprinted on it. We therefore wish to avoid this situation, by creating a synthetic sample of galaxies at the (assumed to be) true redshift under consideration, constructed from as many of the SV1 ELGs as feasible.

Specifically, for a bin at redshift $z_i$, we select all galaxies between $z_i - 0.05$ and $z_i$ and redshift their model spectra to $z_i$, and then compute the Doppler bias as if all of those galaxies were at $z_i$. The width of the bin, 0.05, is chosen to balance considerations of ensuring a sufficiently large sample size, against potential redshift evolution in the ELG samples. Specifically, we assume similar color properties between galaxies in the same redshift bin, which is more likely to be true if the redshift bin is narrower. 
We test whether this assumption is valid by changing the binsize between 0.01 and 0.2. We find that as the binsize increases from 0.05 to 0.15, the average value of $b_D$ in the redshift range of interest, $0.8 < z < 1.0$, decreases by 30\%, but if we decrease the binsize to 0.01, $b_D$ only changes by 10\%. On the other hand, using a binsize much smaller than 0.05 leads to considerably larger errors on $b_D$ due to the small sample sizes; we therefore choose 0.05 for the binsize as the best balance between ensuring a sufficiently large sample and measuring a convergent $b_D$ that is not affected by color evolution across the binsize. We also note that this procedure leads to correlations between neighboring bins (since many of the galaxies are in common) and therefore refrain from quoting errors that require us to combine multiple bins at different $z_i$.

We only consider galaxies at redshifts less than $z_i$ (rather than a symmetric range between $z_i - 0.05/2$ and $z_i + 0.05/2$) because we find that the model spectra can suffer from issues when extrapolated outside of the DESI wavelength coverage of 3600 \AA\ to 9800 \AA.
In particular, when the model spectra are not constrained by data, there is nothing preventing them from dropping below zero. This becomes an issue when shifting the galaxies to lower redshifts, since the upper wavelength coverage of DESI (9800 \r{A}) is slightly less than the upper wavelength of the $z$-band (10000 \r{A}). On the other hand, DESI's lower wavelength of 3600 \AA\ is quite a bit shorter than the 
blue end of the $g$-band (4000 \r{A|}).
The upper cutoff would thus lie within the $z$-band for galaxies with redshift greater than $z_i$ (causing issues with spurious negative flux); but the lower cutoff does not move into the $g$-band for galaxies with redshift greater than $z_i - 0.05$.
Hence, due to this asymmetry between the DESI spectrum coverage and the bandpass locations, we only consider galaxies at $z^{\textrm{meas,j}} < z_i$ when constructing the superset at each $z_i$.

\subsection{Measuring synthetic magnitudes and redshifting them}
\label{sec:synth_mags}
To compute synthetic $grz$ magnitudes for each galaxy, we start with the model DESI spectra, which are given at $z=0$, redshift them to $z^{\textrm{meas},j}$, and convolve with the appropriate filter response curve \cite{speclite}. 
This begins with the filter response convolution operator.
\begin{equation}
    F[S_a,f_\lambda] = \int_0^\infty d\lambda \, f_\lambda(\lambda / ( 1 + z^{\textrm{meas,j}})) S_a(\lambda) \frac{\lambda}{hc},
\end{equation}
where $h$ is Planck's constant and for clarity, we have not set $c$ to 1. Additionally, $f_\lambda(\lambda)$ is the rest-frame model DESI spectrum for each object, corrected as described below, $S_a(\lambda)$ is the filter response curve as shown in Fig.~\ref{fig:filters}, and $a$ indexes the filters $grz$.

To compute the magnitudes, we normalize with the model AB reference source which is used to define the photometric system's zeropoint: $F[S_a,f_{\lambda,0}]$. The reference flux density for the AB system is defined:
\begin{equation}
    f_{\lambda,0}^{AB}(\lambda) = \frac{c}{\lambda^2}(3.631\times10^{-20}\text{erg s$^{-1}$cm$^{-2}$Hz$^{-1}$}).
\end{equation}
Normalizing the filter response convolution operator and simplifying, we obtain
\begin{equation}
    \frac{F[S_a,f_\lambda]}{F[S_a,f_{\lambda,0}]} = \frac{\int d\lambda \, f_\lambda(\lambda / ( 1 + z^{\textrm{meas,j}})) S_a(\lambda) \lambda}{(3.631\times10^{-20}\text{erg s$^{-1}$cm$^{-2}$Hz$^{-1}$})\int d\lambda \, S_a(\lambda) c/\lambda}.
\label{eqn:synth_mag}
\end{equation}
We then convert flux to AB magnitude
\begin{equation}
    m_a = -2.5\log_{10}\qty(\frac{F[S_a,f_\lambda]}{F[S_a,f_{\lambda,0}]}) = -2.5 \log_{10}(\langle f_{\nu}\rangle_a) - 48.6,
\label{eqn:synth_mag_def}
\end{equation}
where
\begin{equation}
    \langle f_{\nu}\rangle_a = \frac{\int d\lambda \, f_\lambda(\lambda / ( 1 + z^{\textrm{meas,j}})) S_a(\lambda) \lambda}{\int d\lambda \, S_a(\lambda) c/\lambda}.
\end{equation}

We apply a small correction to the model DESI spectra to remove unphysical negative emission lines. These arise because
\texttt{Redrock} does not place any physical priors on the emission and absorption
lines. One particularly common failure mode in the ELG spectra is [OIII] 4959 and 5007 \AA\ absorption rather than emission, likely tied to sky subtraction residuals from OH skylines at the red end of the spectrum. This can create strongly negative flux in the [OIII] region, leading to a negative overall flux or spurious flux shifts related to the redshifting of these unphysical absorption lines.
We correct models with unphysical absorption by fitting a line to the emission spectrum values in the ranges of $4910(1+z)$ to $4935(1+z)$ \AA\ and $5060(1+z)$ to $5084(1+z)$ \AA\ on either side of the lines. Then we replace the points in the spectrum with the linear fit in these ranges if the points are more than two standard deviations below the linear fit. 

After computing $m_a$, we have two measurements for the broad-band photometry for each galaxy: the synthetic magnitudes from Eq.~\ref{eqn:synth_mag}, and the broad-band fluxes from the Legacy Survey DR9 imaging survey.
In theory, these two measurements should be exactly consistent with each other;\footnote{The synthetic magnitude should agree exactly with the fiber magnitude, since the fiber magnitude is measured within an aperture that is the same size as a DESI fiber; the total magnitude will be different since it captures more light.} however, in practice, due to observational errors (i.e.\ Poisson noise on the measured fluxes and spectra), this is not the case.
Since the DESI target selection uses the broad-band fluxes from the imaging, we start with the DR9 fluxes, and only use the synthetic magnitudes to measure the \textit{shifts} in magnitude due to peculiar velocity. That is, for each galaxy indexed by $j$, when applying a peculiar velocity of $\Delta z$, we compute
\begin{equation}
    \mathcal{S}_{a,\textrm{z only}}^j(z, z + \Delta z) = m_a^j(z + \Delta z) - m_a^j(z)
\end{equation}
using the synthetic magnitudes from Eq.~\ref{eqn:synth_mag_def};
or equivalently, writing the shift function $\mathcal{S}$ in terms of fluxes
\begin{equation}
    \mathcal{S}_{a,\textrm{z only}}^j := f_\lambda(\lambda / (1 + z)) \rightarrow f_\lambda(\lambda / (1 + z + \Delta z))
\end{equation}
where $\mathcal{S}_{a,\textrm{z only}}^j$ is the function that modifies the magnitudes of the $j$th galaxy corresponding to a redshift from $z$ to $z + \Delta z$. The notation above is defining $\mathcal{S}$ as the operation that shifts $f_\lambda(\lambda / (1 + z))$ by $\Delta z$, the operation indicated with the arrow.
When applying the small redshift of $\Delta z$, we do \textit{not} also modify the flux to account for the increased luminosity distance, since changes in galaxy brightness are treated separately
in Eq.~\ref{eqn:delta_g_relativistic} and are related to the magnification bias $s$.
We indicate this with the subscript ``z only.''

We also modify the observed fluxes to preserve the rest-frame $f_\nu$, when shifting from $z^{\textrm{meas,j}}$ to $z_i$. 
Since we are setting up the ELG superset, in this step we \textit{do} scale the observed flux with the luminosity distance
\begin{equation}
    \mathcal{S}_{\textrm{$z + D_L$ shift}}^j := f_\lambda(\lambda / (1 + z^{\textrm{meas,j}}))) \rightarrow  f_\lambda(\lambda / (1 + z_i)) \frac{D_L(z^{\textrm{meas,j}})^2}{D_L(z_i)^2} \frac{1 + z^{\textrm{meas,j}}}{1+z_i}
\end{equation}
As described above, we consider an asymmetric set of galaxies, redshifting those within ($z_i -0.05$, $z_i$) to $z_i$.
As a result, the galaxies are always fainter when moved to $z_i$, since they are all moved further away. The effect of the Doppler bias on galaxy selection depends both on the impact of the peculiar velocity on the spectrum, and the underlying distribution of galaxies as a function of their rest-frame spectrum. In particular, it cares about the distribution of galaxies close to the flux cuts, and if the galaxies are all made fainter, we will be probing a different set of galaxies than we otherwise would have.
This could cause the synthetic sample to measure a different, incorrect, Doppler bias from the true Doppler bias of the underlying ELG sample.
To correct for this, we \textit{brighten} all galaxies by an amount corresponding to a shift of 
half of the redshift binsize used to construct the superset, 0.025 in the fiducial case. Therefore, the final shifted magnitudes for each galaxy are given by:
\begin{equation}
   m_a^{\textrm{shifted,j}} =  m_a^{\textrm{LS,meas,j}} + \mathcal{S}^j_{\textrm{$z + D_L$ shift$- D_L(\Delta z = \textrm{binsize}/2)$ shift}}( z_{\textrm{meas,j}}, z_i) + \mathcal{S}^j_{\textrm{z only}}(z_i, z_i + \Delta z)
\end{equation}
where the subscript ``$-D_L(\Delta z = \textrm{binsize}/2)$ shift''
indicates the brightening described in the previous paragraph. However, we find that in practice, whether or not we apply the brightening of half of the binsize makes very little difference in the Doppler bias, likely because the ELG selection is mainly determined by the color cuts rather than the $g$-magnitude cuts.

We compute the number of galaxies passing the ELG selection cut for the shifted magnitudes, and also for unshifted magnitudes
\begin{equation}
  m_a^{\textrm{unshifted,j}} =  m_a^{\textrm{LS,meas,j}} + \mathcal{S}^j_{\textrm{$z + D_L$ shift$- D_L(\Delta z = \textrm{binsize}/2)$ shift}}( z_{\textrm{meas,j}}, z_i)
\end{equation}

An important assumption in this approach is that the superset is sufficiently complete, i.e.\ that it
contains all galaxies that could have scattered into the underlying ELG
selection due to peculiar velocities.
This means that the superset must have a buffer around the DESI ELG selection, with width equal to the typical shifts in $g-r$ and $r-z$ colors.
The typical shifts in these colors are very small, $\sim$0.005 mags, with the largest shifts reaching up to $\sim$0.2 mags.
As shown in Fig.~\ref{fig:filters}, the SV1 ELG cuts are nearly always sufficiently far away,
except in the region around $r-z \sim 0.15$, $g-r \sim 0.3$, where the two selections cross. This region in color space is mostly populated by higher-redshift ELGs (Fig.\ 4 in \cite{Raichoor23}), and is sparsely populated at $z < 1.2$. 
We test for the effects of incompleteness by shifting the second color cut in Eq.~\ref{eqn:elg_selection} by 0.05 magnitudes, since $b_D$ at $0.8 < z < 1$ is determined by galaxies shifting across this color cut.
As we discuss in Section~\ref{sec:doppler_bias_measurement} below, we find that $b_D$ is somewhat sensitive to incompleteness.

\subsection{Measuring the change in number of ELGs}
\label{sect:measuring_delta_elg}

After the magnitudes are shifted, we count the number of galaxies within the DESI ELG selection cut before and after the peculiar velocity shift. 
We conduct a convergence test along similar lines as that described in \cite{wenzl2023magnification}.
We start by computing two-sided derivatives, i.e.\ measuring
\begin{equation}
    \Delta N(\Delta z_{\textrm{pec}}) = N_{\textrm{color}}(z + \Delta z_{\textrm{pec}}) - N_{\textrm{color}}(z - \Delta z_{\textrm{pec}})
\end{equation}
\begin{equation}
    b_D = -(1+z) \frac{\Delta N(\Delta z_{\textrm{pec}})}{2 \Delta z_{\textrm{pec}} N_{\textrm{color}}(z)}
\end{equation}
Determining $b_D$ in this way has errors that are quadratic in the step-size $\Delta z$ \cite{wenzl2023magnification,Samuroff19}. Thus, we measure $\Delta N$ for several different step sizes.
We then create independent samples by differencing $\Delta N$ between sequentially larger step sizes, measuring $\hat{b}_D$ in each step, and fitting a quadratic as a function of increasing step size.
If $\Delta z_{\textrm{pec}}$ is small compared to the scales
on which $b_D$ varies, the quadratic will be a good fit, and evaluating the quadratic at a stepsize of zero will be the optimal estimator for $b_D$.
We can perform this test at each redshift, $z_i$ under consideration on our grid.
We test values of $z_{\textrm{pec}}$ ranging between 0.001 and 0.005 (i.e.\ stepsizes between 0.002 and 0.01). We find
that at some $z_i$, the quadratic is a good fit, but at others, it is not.\footnote{A further subtlety arises from the likelihood that is minimized when the quadratic is fit. In the low-$N$ limit, it becomes important to consider that we are differencing two Poisson distributions. Thus, the likelihood is not necessarily Gaussian. We find that some of the bad fits are in the low $N$ limit, but some of them are also at very large $N$ where the central limit theorem guarantees Gaussianity, implying that the poor fits are due to intrinsic variability in $b_D$ as a function of stepsize, not non-Gaussianity of the fitted likelihood.}
This points to sharp changes in $b_D$ as a function of the step-size, which is not surprising given the sharpness of the spectral features under consideration (most prominently [OII] emission, but also other features as well). This means that the fit should contain higher even powers of $\Delta z$, which can have large coefficients making them larger than the $\Delta z^2$ term.
We could push to even smaller values of $\Delta z_{\textrm{pec}}$; however, the resolution of the filter bandpasses, $S_a(\lambda)$, corresponds to $\Delta z_{\textrm{pec}} \sim 0.027$ for [OII] redshifted to $z = 1$, and thus a smaller $\Delta z_{\textrm{pec}}$ will unavoidably miss even finer features in $b_D$ that come from features in the filter curves that are simply not measured at the resolution that is available.
Balancing these considerations, the best procedure is to use a two-sided derivative with $\Delta z_{\textrm{pec}} = 0.001$, which is the smallest step-size resolvable given the filter resolution, and avoids extrapolating to smaller step-sizes assuming the validity of the quadratic fit. The plot illustrating this test is shown on the right hand side of Fig. \ref{fig:bd}, and shows that different choices for the step-size make very little difference.

\subsection{Doppler bias measurement}
\label{sec:doppler_bias_measurement}

\begin{figure}
    \centering
    \includegraphics[width=\textwidth]{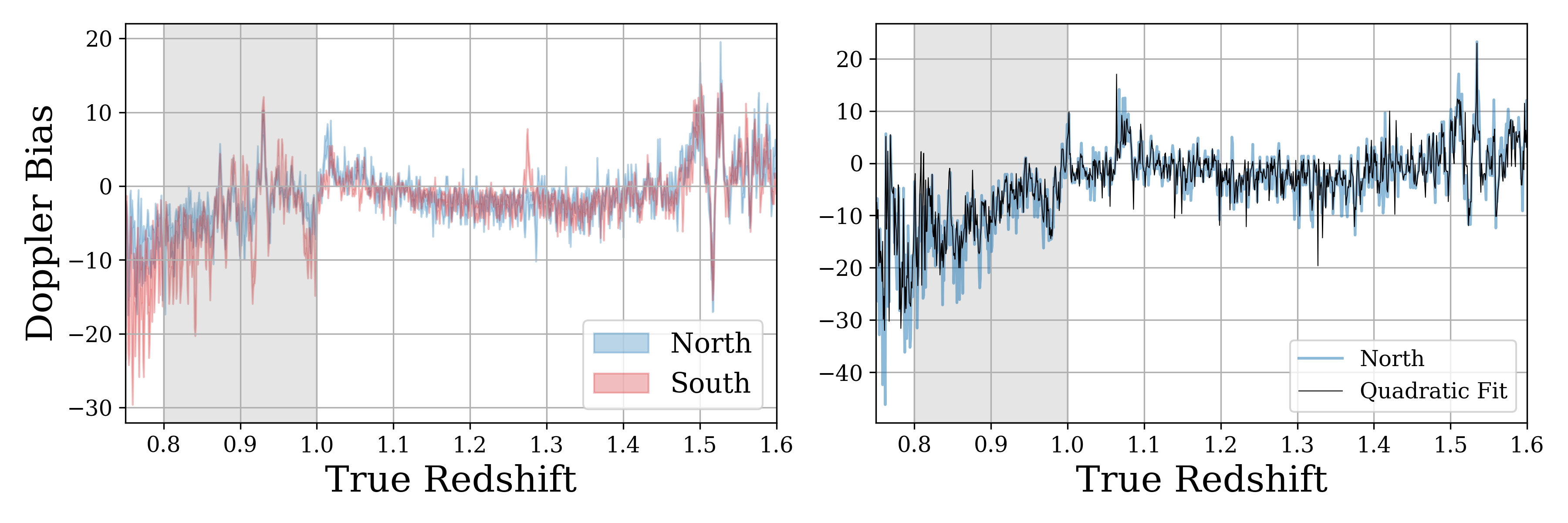}
    \caption{\textit{Left:} Doppler bias for the North and South samples, with the 1$\sigma$ uncertainty band given by the shaded region.
    We emphasize that the Doppler bias is computed as a function of the redshift, $z_i$, of the ELGs, and to determine its impact on observables, it must be averaged across a bin in observed redshift.
    The grey rectangle shows the 0.8--1.0 redshift range that we study further since it has the largest $b_D$ impact.
    \textit{Right:} Comparison between two methods for estimating the Doppler bias: the smallest step size considered of $\Delta z = 0.002$ (solid; fiducial method), and a quadratic fit to Doppler bias measured in step sizes between $\Delta z = 0.002$ and 0.01 (dashed), to ensure that the  results are independent of step size. The two results are nearly indistinguishable.}
    \label{fig:bd}
\end{figure}

We show the estimated $b_D$ in Fig.~\ref{fig:bd} between the North and the South. On the right panel, we compare the results from the quadratic fit to those from the smallest step-size considered of $\Delta z_{\textrm{pec}} = 0.001$. The results are nearly indistinguishable; as a result, our results are robust to convergence issues caused by a too-large $\Delta z_{\textrm{pec}}$.

It is important to note that Fig.~\ref{fig:bd} is not an observable quantity, but rather gives the Doppler bias as a function of redshift $z_i$ in our constructed ELG supersets. Determining the impact on observables (i.e.\ galaxy power spectra) requires averaging $b_D$ within some pre-defined bin in \textit{observed} redshift.

The Doppler bias shows many narrow features, corresponding to specific redshifts
where strong emission lines redshift across features in the bandpasses. If the bandpasses were simply tophats, these redshifts would only occur were emission lines intersect the edges of the bandpasses; however, since the filter shapes are more complex, narrow features can exist in the Doppler bias at other redshifts too.
The $b_D$ features are slightly more prominent in the South filters, mostly due to the sharper drop at the red end of the South $r$-band, and the notch in the South $r$-band at 6900 \AA\ (Fig.~\ref{fig:filters}).

\begin{figure}
    \includegraphics[width=\textwidth]{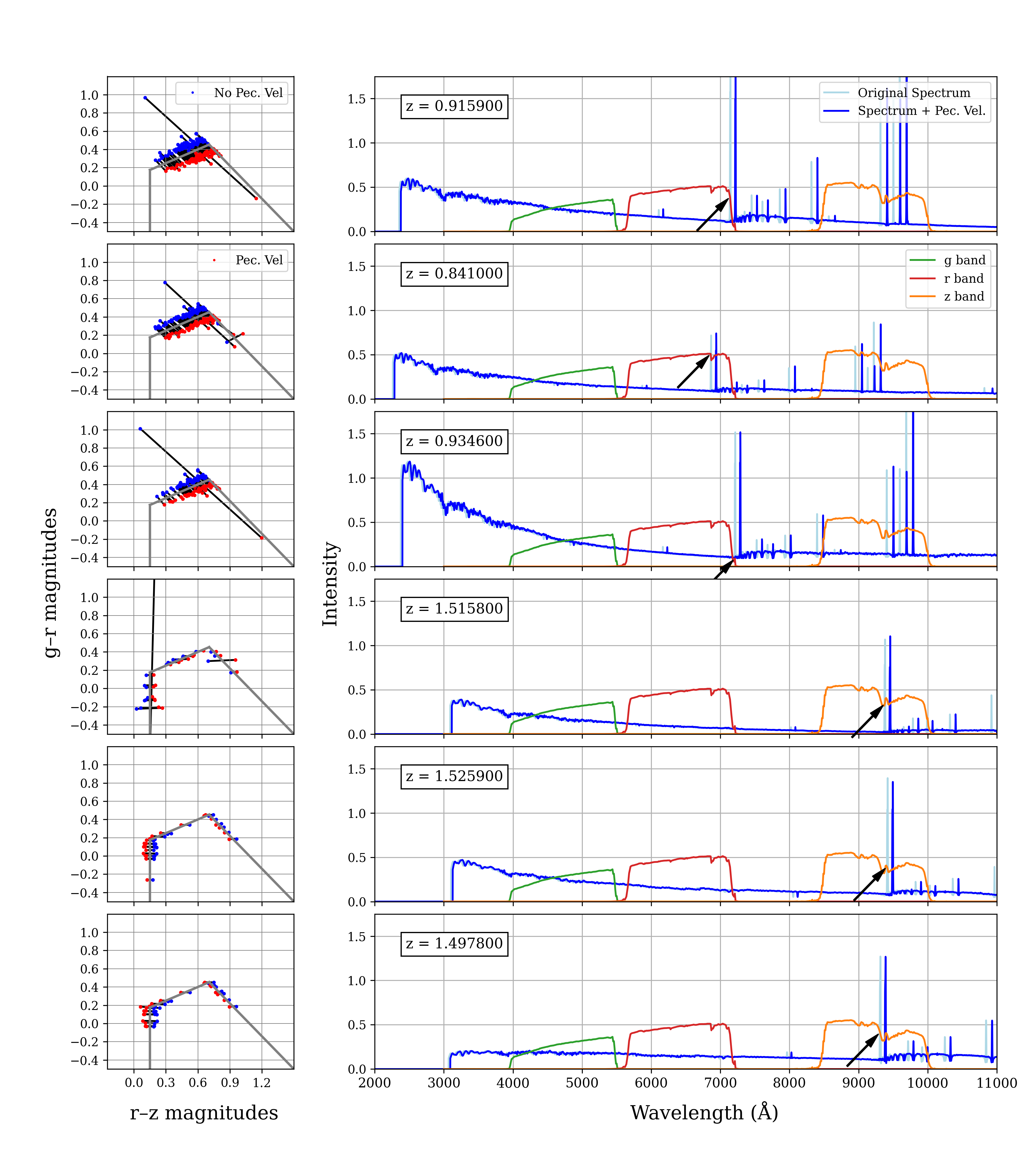}
    \caption{\textit{Left:} ELG $g-r$ and $r-z$ color shifts. Each point shows an ELG at redshift $z_i$, comparing measured (blue) and velocity-shifted (red) photometry. Only ELGs that cross the DESI ELG selection box (grey) when shifted with small peculiar velocity are shown. Shifts are scaled by a factor of 10 for clarity. \textit{Right:} Example spectrum at $z_i$ with South $grz$ filters, comparing original (light blue) and velocity-shifted (dark blue) spectra, shifted by an exaggerated $z=0.02$ for visibility. Arrows highlight where the [OII] line aligns with a sharp filter drop, causing magnitude shifts in the color-color plots.
    \label{fig:bd_features}}
\end{figure}

In Fig.~\ref{fig:bd_features}, we explore some specific redshifts and plot some example spectra to explain where the trends in Fig.~\ref{fig:bd} are coming from. Fig.~\ref{fig:bd_features} shows the origin of some of the most prominent features in Fig.~\ref{fig:bd}. There are several features at $z < 1$ coming from [OII] shifting across features in the $r$-band filter that decrease the $r$ magnitude. Peculiar velocity shifted galaxies are thus fainter in the $r$-band, which makes $g-r$ smaller and $r-z$ larger, thus shifting galaxies down and to the right in the color-color plot. This causes many galaxies to scatter into the ELG color selection, resulting in a positive derivative of $N_{\textrm{color}}$ and thus a negative Doppler bias. These features are prominent at $z = 0.8410$ ([OII] passing over the notch in the South $r$-band at 6900 \AA\,); $z = 0.9159$ from [OII] passing out of the $r$-band at 7100 \AA\,; and a rise at $z=0.9303$ and a dip at $z = 0.9386$ from [OII] passing over a small rise and dip in $r$-band at 7200 \AA\, just after the  steep dropoff of the $r$-band at 7100 \AA.

There are also a number of features arising from emission lines passing into and out of the $z$-band. The beginning of $z$-band at 8500 \AA\ is more gradual than the end of $r$-band at 7100 \AA\,; hence the $b_D$ rise at $z = 1.276$ is less prominent than the dip at $z = 0.9159$, and only present in the South where the $z$-band filter rises more steeply. The most prominent features caused by $z$-band are the rises and dips at $z = 1.4979$, 1.5158, and 1.5260. These come from [OII] passing over some large features in the middle of the $z$-band filter, starting around 9350 \AA\,; these features are present in both the North and South $z$-band filters and are thus present in both the North and South $b_D$ curves. Other features come from the [OIII] emission lines at 4959 and 5007 \AA\,.
The dips at $z \sim 0.99$ come from these lines passing out of the $z$-band at 10000 \AA\,. The dips at $z = 0.8605$ and $0.8820$ come from the more prominent of these lines, [OIII] 5007 \AA\,, passing over the dips in the middle of $z$-band at 9350 and 9450 \AA\,.

In addition to the narrow features created by the lines, there is also a broad trend for slightly negative $b_D$ at $z < 1$, and nearly zero $b_D$ at $z > 1$.
This is likely coming from the broad Balmer break at 3700--4000 \AA\ passing out of the $r$-band, creating a similar negative $b_D$ feature to the one created by [OII] passing out of the $r$-band, but extended over a much larger region because the Balmer rise is much more gradual than the [OII] line.
We single out the redshift range 0.8 to 1.0, which includes several of the largest narrow spikes, as well as the broad-band feature of negative $b_D$ coming from the Balmer break. 
Since this is the redshift range where $b_D$ is largest, we study its impact on clustering statistics in this range by considering the clustering of a bin of ELGs at these redshifts. 
As emphasized in Fig.~\ref{fig:flowchart} above, this requires us to translate from $b_D$  as a function of redshift, $z_i$, to an observable quantity that gives the impact of the Doppler bias in the clustering in a particular redshift bin. To average $b_D$ within the bin, we compute the corresponding $\Delta N/N$ at each $z_i$, multiply by the comoving number density $\bar{n}(z_i)$, sum $\Delta N$, and then finally divide by the summed comoving number density. 
We find an averaged $b_D$ of $-5.9$ for the South sample and $4.7$ for the North.
We then use this value as input to GaPSE in Section~\ref{two-point stats}.
There are larger features in $b_D$ around $z \sim 1.5$ than around $z \sim 0.9$. However, since they are both positive and negative, when averaging in a bin from  $1.4 < z < 1.6$, we find that $b_D$ is $\sim 3 \times$ smaller than at $0.8 < z < 1.0$. Furthermore, we lack another high-density tracer at $1.4 < z < 1.6$, unlike at $0.8 < z < 1$ where the LRGs are convenient for cross-correlation. As a result, we focus on $0.8 < z < 1$ where the effects of $b_D$ are largest and most easily detectable.
As discussed in Sec.~\ref{sec:synth_mags}, our results could be affected by incompleteness in the parent ELG sample at  $r-z \sim 0.15$, $g-r \sim 0.3$. The density of $z \sim 0.9$ galaxies is relatively low in this region in color-space (where the average redshift is typically higher), so incompleteness should only have a modest impact on $b_D$ at these redshifts.
However, if we are missing galaxies from the superset, this would cause us to under-estimate $b_D$ at $z \sim 0.9$, where peculiar velocities are causing galaxies to enter the ELG sample by making them fainter in $g$ magnitude. If the superset is incomplete, then even more galaxies should scatter into the sample than we observe, leading to a larger, even more negative $b_D$. We find that our results are rather sensitive to the selection cut in $g-r$ and $r-z$, with a 0.05 change in magnitude leading to $\Delta b_D \sim 3$ (although the broad trends and positions of the narrow spikes are qualitatively similar). This suggests that
our results are sensitive to incompleteness in the parent sample. Future work with more extended parent samples may therefore be necessary to more accurately measure $b_D$.

\section{Linear bias, magnification bias and evolution bias}
\label{bias measuring}
\subsection{Linear bias}
We estimate the linear bias $b_1$ from the projected correlation function $w_p$ of the SV3 ELGs at $20 < r_p < 80$ $h^{-1}$ Mpc, comparing the measured $w_p$ to the expectation from the HALOFIT nonlinear power spectrum \cite{Mead20} as computed by CAMB in the fiducial cosmology \cite{Lewis1999,Howlett2012}. 
We use the fiducial weighting scheme provided in the DESI EDR clustering catalogs, which includes FKP and completeness weights but does not correct for imaging systematics due to the small area of the DESI SV3 survey, making measurement of the trends difficult, and the fact that they mainly affect larger scales than those under consideration here \cite{DESI23b}.
As we see in Table \ref{tab:biases_table}, the linear bias is approximately constant until $z = 1.3$, and increases thereafter. 

Since the peculiar velocity terms are enhanced in the multi-tracer dipole proportional to the difference in linear biases, we explored the prospect for splitting the ELG sample into multiple subsets.
We measured ELG clustering in bins split by $g$ magnitude, but found that the linear bias of the different ELG bins was very similar.
Other splits in ELG properties may yield a larger bias difference \cite[e.g.,][]{Hagen24}, but we do not consider them here. Instead, we considered the cross-correlation between ELGs and LRGs, which overlap the ELGs well, especially at $z \sim 0.9$ where the Doppler bias is largest, and gives us a large sample of galaxies with considerably higher linear bias.
 
The LRG sample has been well-characterized in past work (e.g.,\ tomographic CMB lensing cross-correlation) and we therefore use $b = 2$ and $s = 1$ \cite{White21,Zhou23}. We compute the evolution bias from the measured number density $\bar{n}(z)$ in the same way as for the ELGs.
LRG target selection uses the infrared WISE W1 and W2 bands (3.4 and 4.6 $\mu$m) in addition to $grz$. We do not have model spectra at these wavelengths, so we cannot calculate the Doppler bias for the LRGs. However, since the LRGs are dominated by continuum emission rather than emission lines, we expect the Doppler bias to be negligible, and therefore set it equal to zero in the power spectrum calculation.

\begin{table}[]
    \centering
    \begin{tabular}{|p{3cm}|p{3cm}|p{1cm}|}
        \hline
        Redshift  & $b_1$ & $\chi^2$ \\
        \hline
        0.85 & 1.25 $\pm$ 0.08 & 6.3 \\
        0.95 & 1.17 $\pm$ 0.07 & 5.0 \\
        1.05 & 1.29 $\pm$ 0.09 & 13.9 \\
        1.15 & 1.23 $\pm$ 0.08 & 0.2 \\
        1.25 & 1.30 $\pm$ 0.09 & 3.4\\
        1.35 & 1.59 $\pm$ 0.10 & 1.4\\
        1.45 & 1.72 $\pm$ 0.10 & 1.8\\
        1.55 & 1.41 $\pm$ 0.14 & 0.7 \\
        \hline
    \end{tabular}
    \caption{Linear bias for the ELG sample, measured from the projected correlation function at $20 < r_p < 80$ $h^{-1}$ Mpc with 5 degrees of freedom.}
    \label{tab:biases_table}
\end{table}

\subsection{Magnification bias}
\label{magnification bias}

\begin{figure}
    \centering
    \includegraphics[width=\textwidth]{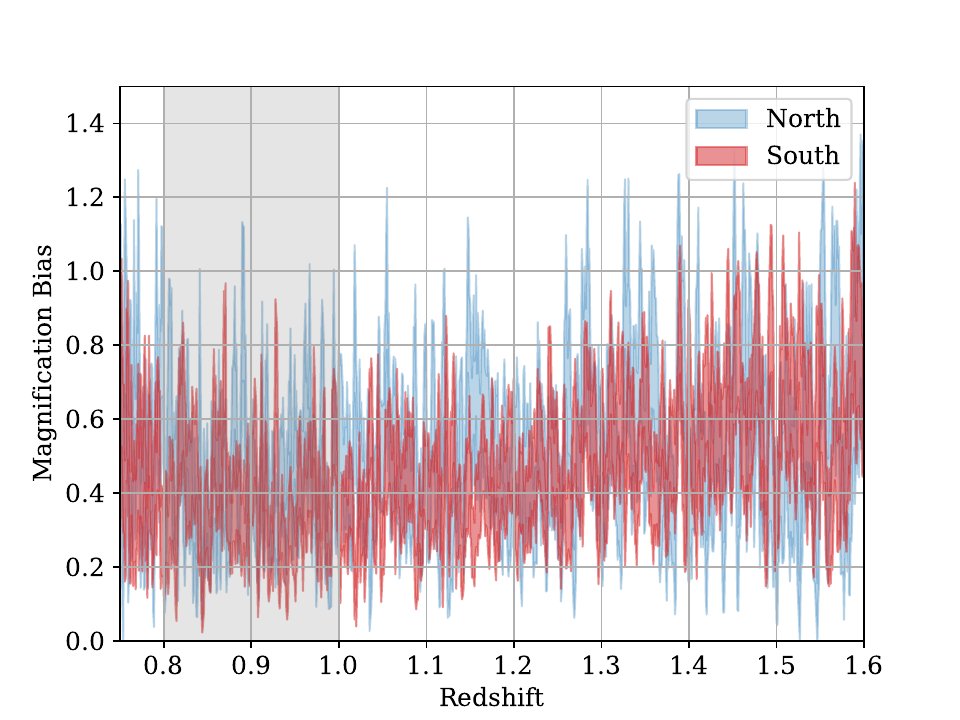}
    \caption{Magnification bias for North and South galaxy samples, showing the 1$\sigma$ uncertainty band (shaded). The grey rectangle shows the 0.8--1.0 redshift range that we study further since it has the largest $b_D$ impact.}
    \label{fig:mag_bias}
\end{figure}
The magnification bias is sensitive to the overall distribution of $g_{\textrm{fib}}$ and $g$ magnitudes, since lensing magnification is achromatic and leaves the $g-r$ and $r-z$ colors unchanged.
As shown in Eq.~\ref{eqn:elg_selection}, there is no ELG selection cut that couples together colors and magnitudes; hence, the only changes that will matter for magnification are those that affect the upper and lower flux limits, $g > 20$ and $g_{\textrm{fib}} < 24.1$.

After creating the synthetic sample, measuring magnification bias is straightforward. Since we have the parent sample of galaxies, we can simply magnify the sample by a certain amount and measure the change in number, unlike the two-direction shifting described in \cite{wenzl2023magnification,ElvinPoole23} to measure magnification bias in the case where only the final targeted sample is available. We therefore add $\Delta m = 0.01$ to each magnitude, and compute the magnification bias by taking the finite difference of the number of galaxies at the shifted magnitude, and the original number of galaxies.

We only shift the galaxy magnitudes fainter (increasing the magnitudes) because the faint end of the ELG sample is close to the detection limit in the North imaging ($g=24.1$ at the 5$\sigma$ limit for a 0.45 arcsec exponential profile in $g$-band, from Table 1 in \cite{Raichoor23}). Shifting the magnitudes brighter (decreasing the magnitudes) would mean that magnification would add galaxies to the sample that were not detected in the original Legacy Survey imaging. Since these galaxies are by definition not in the catalog, our measurement of magnification bias would therefore be biased.

A magnification by 0.01 magnitudes does not increase fiber magnitudes by the same amount for all types of galaxies.
Since lensing magnification makes galaxies bigger but preserves surface brightness, fixed-aperture
magnitudes cannot receive the entire increase in flux, since part of the magnified galaxy is moved outside the aperture. Following \cite{Zhou23}, we assume that the total fluxes receive all of the change in flux from magnification, but the impact on fiber magnitudes is reduced.
Following Appendix C in \cite{Zhou23}, we translate the change in total magnitudes to the change in fiber magnitudes.
Of the faintest 0.01 magnitudes of main selection ELGs, we find that 29.6\% are point sources, 59.6\% are round exponential, 8.8\% are exponential, 2.0\% are deVaucouleurs and 0\% are Sersic.
We interpolate the fiber flux divided by the total flux as a function of half-light radius (or half-light radius and aspect ratio for de Vaucouleurs galaxies), and then weight by the observed
distribution of half-light radius and aspect ratio.
Weighting by the distribution of each type, we find a response of 0.8 (i.e.\ magnification of the flux by a factor $F$ increases the fiber flux by $0.8F$), which is constant with redshift.
The response is larger than that found by \cite{Zhou23} for DESI LRGs, because the faint end ELGs are fainter and more point-source-like.

We show the magnification bias as a function of the redshift in Fig.~\ref{fig:mag_bias}. Averaging between 0.8 and 1.0 (again weighting by the comoving number density), we find $s = 0.44$ for North and $s = 0.38$ for South. Our values for magnification bias are lower than those of \cite{Karim24}, who find $s = 0.89$, also for DESI ELGs. However, they are using the ELG target sample combining both \texttt{ELG\_LOP} and \texttt{ELG\_VLO} selections, rather than the spectroscopically confirmed ELGs from \texttt{ELG\_LOP} that we use.
Moreover, they apply some additional cuts to improve the purity of the target sample (intended to remove low-redshift interlopers, quasars, and faint or high-redshift ELGs where a good redshift cannot be obtained).
As a result, their sample selection is somewhat different from ours, and it is thus unsurprising that they find a different value of the magnification bias.

\subsection{Evolution bias}
\label{evolution bias}
\begin{figure}
    \centering
    \includegraphics[width = \textwidth]{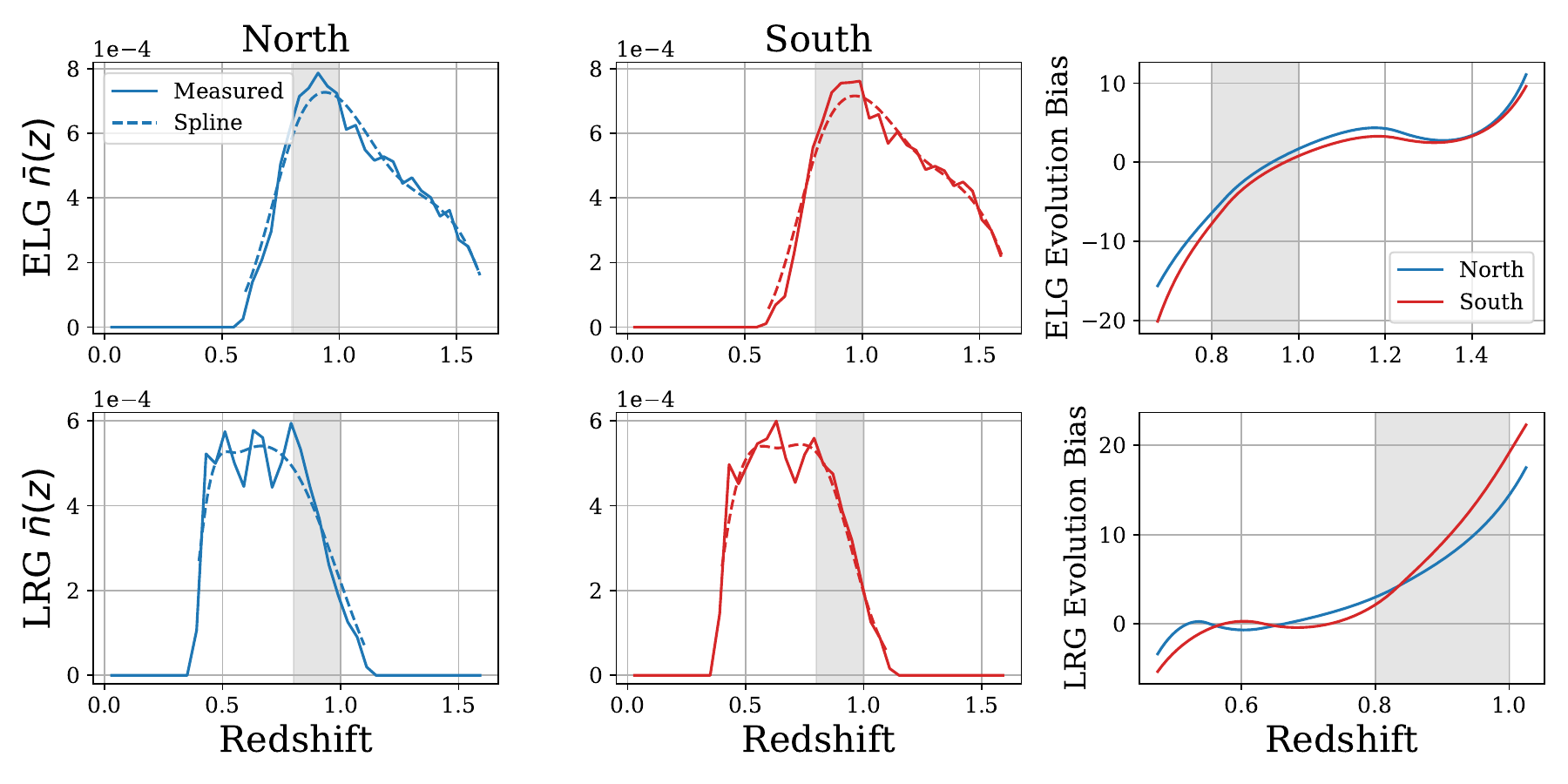}
    \caption{The left and middle columns show the comoving number density of the ELG and LRG samples as a function of redshift. The left column is from the North galaxy sample while the middle column is from the South galaxy sample. The right column shows the evolution bias for both the ELG and LRG samples computed from $\bar{n}(z)$ using Eq.~(\ref{eqn:ev_bias}). The grey rectangle highlights the redshift range of interest.}
    \label{fig:galvsredsh}
\end{figure}

The evolution bias is computed using the comoving number density of galaxies as shown in Fig. \ref{fig:galvsredsh}. 
We use the measured $\bar{n}$ from SV3 tabulated by the DESI survey.\footnote{\url{https://data.desi.lbl.gov/doc/releases/edr/vac/lss/}} Subsequently, we smooth the noisy measured $\bar{n}(z)$ using a smoothing spline. We verify that the fluctuations seen in $\bar{n}(z)$ are driven by cosmic variance due to the small area covered, by comparing to the redshift distributions from the first two months of the DESI main survey as shown in Fig.~19 in \cite{Raichoor23}. The fluctuations in $\bar{n}(z)$ are smaller in this sample, as would be expected since it covers a much larger area.
Averaging over the $0.8 < z < 1.0$ range and weighting by $\bar{n}(z)$, we find $-1.2$ and $-2.1$ for the North and South ELG sample and 8.1 and 10.2 for the North and South LRG sample.

\begin{figure}
    \centering
    \includegraphics[width = \textwidth]{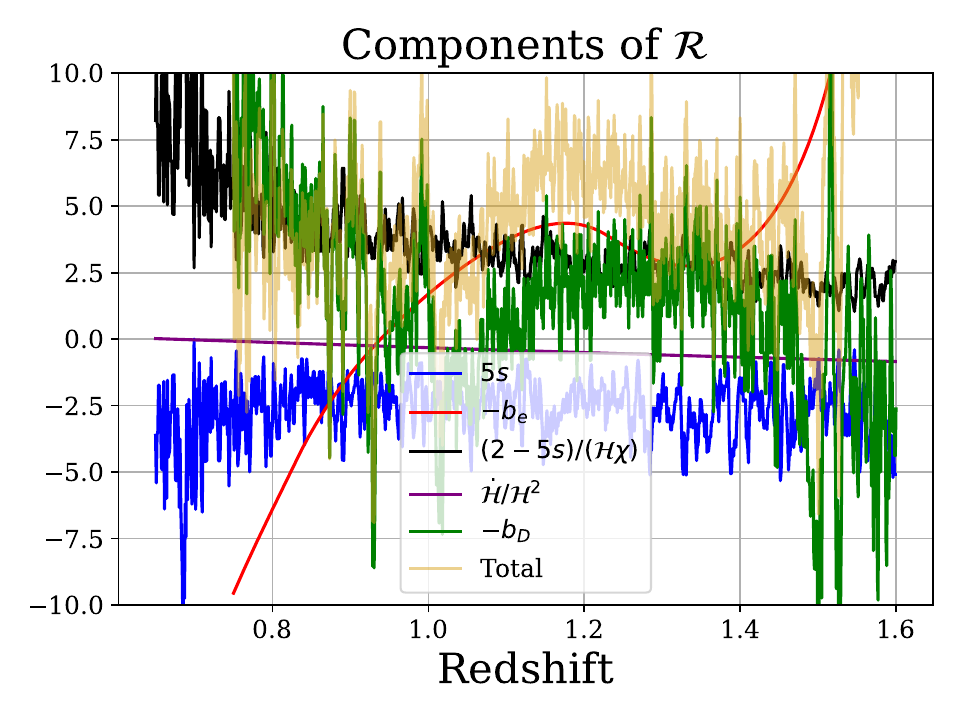}
    \caption{Components of $\mathcal{R}_v$, which quantifies the line-of-sight peculiar velocity modulation of galaxy density, in Eq.~\ref{eqn:b_D_correction}. The vertical axis is unlabeled because each of these components is unitless.}
    \label{fig:contributions_to_r}
\end{figure}

To summarize the contributions of the various terms that multiply the velocity in Eq.~\ref{eqn:delta_g_relativistic} (which are grouped together in $\mathcal{R}_v$ as defined in Eq.~\ref{eqn:R}), we plot them together in Fig.~\ref{fig:contributions_to_r}. This shows that they are generally of similar magnitude, and there are significant cancellations between positive and negative terms.

\section{Impact on two-point statistics}
\label{two-point stats}

\begin{figure}
    \centering
    \includegraphics[width=\textwidth]{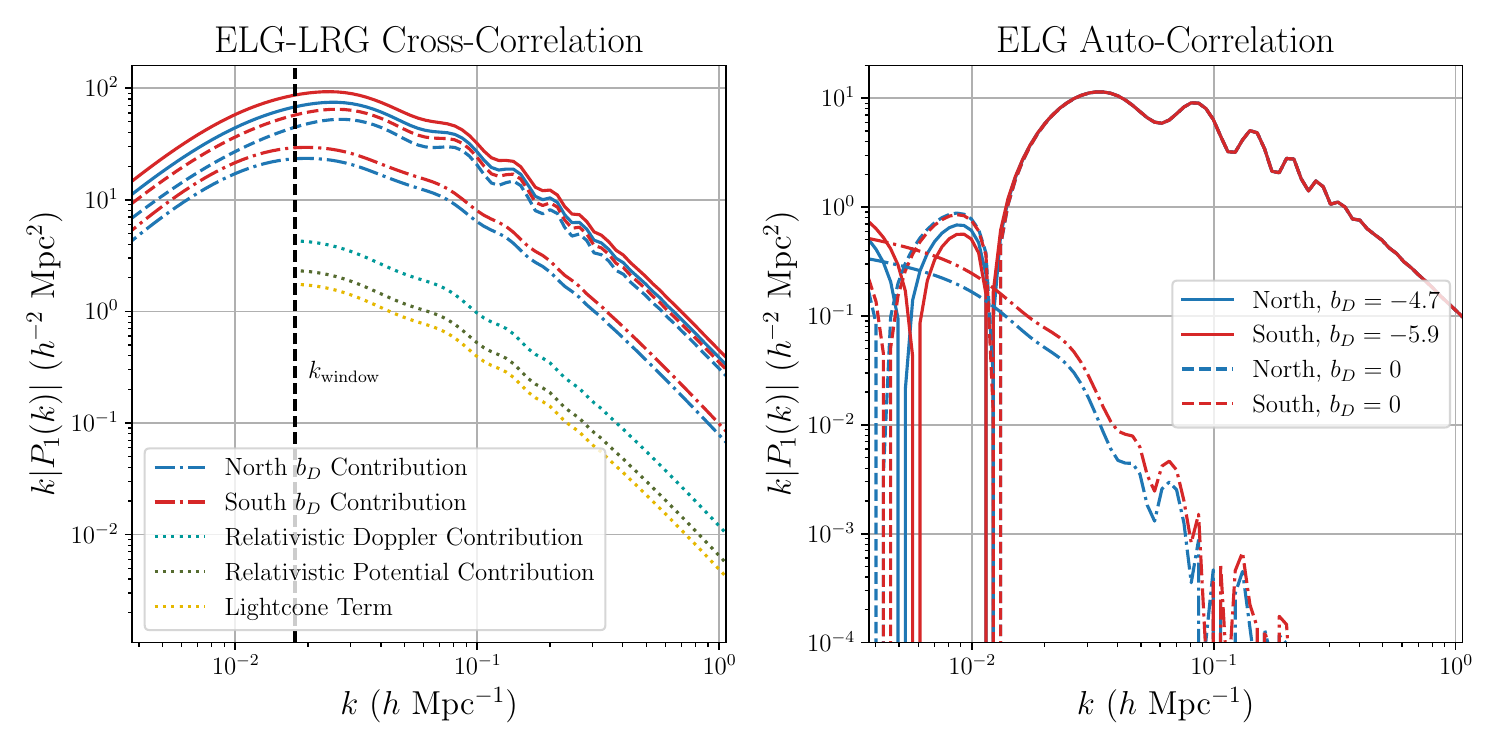}
    \caption{Dipole of the power spectrum with and without Doppler bias, using $b_D = -5.9$ (South) and $-4.7$ (North).
    The left panel shows the ELG-LRG cross-correlation, and the right panel shows the ELG autocorrelation, in both cases at $0.8 < z < 1$. 
    We also compare the cross-correlation dipole to the relativistic Doppler, relativistic potential, and lightcone terms, following the nomenclature of Fig.\ 1 in \cite{BeutlerDiDio20}. These terms were computed from the linear power spectrum without a window function, leading to a deviation in shape on large scales ($k < k_{\textrm{window}}$, the dashed vertical line) vs.\ the window-convolved GaPSE calculations; we thus do not show them at $k < k_{\textrm{window}}.$
    We used the biases from Table \ref{tab:all_biases}, computed as described in Sections \ref{doppler bias} and \ref{bias measuring}.} 
    \label{fig:two_pt_stats}
\end{figure}

In Fig.~\ref{fig:two_pt_stats} we show both the ELG auto-correlation and the ELG-LRG cross-correlation, in the redshift range of interest at $0.8 < z < 1.0$. We compute the dipole both with and without the Doppler bias.
As expected, the impact of $b_D$ on the ELG-LRG cross-correlation is much larger.
Adding $b_D$ only affects the ELG auto-correlation on very large scales, at $k < 0.03$ $h$ Mpc$^{-1}$.
The impact of $b_D$ on the cross-correlation is also scale-dependent, increasing the cross-correlation by 50\% at $k < 0.1$ $h$ Mpc$^{-1}$ and 25--30\% at $k > 0.3$ $h$ Mpc$^{-1}$. 
We also find a small impact of $b_D$ on large-scales in the monopole $P_0$ and quadrupole $P_2$: for the ELG auto-correlation, $b_D$ shifts the monopole by 0.5\% at $k = 0.01$ $h$ Mpc$^{-1}$ and the quadrupole by $\sim$2\%, whereas for the ELG-LRG cross-correlation, $b_D$ changes the monopole by 0.1\% at $k = 0.01$ $h$ Mpc$^{-1}$ and the quadrupole by 2\%.
Due to the scale-dependence of the change, it is unlikely to affect cosmological parameter inference (i.e.\ measurements of $f\sigma_8$), but it could potentially affect $f_{\textrm{NL}}$ constraints. However, the impact of $b_D$ on the cross-correlation dipole is considerably larger, and we will hereafter only consider its impact on $P_1(k)$, leaving its impact on the monopole and quadrupole to future work.

To provide further physical context to the impact of the Doppler bias on the cross-correlation dipole, we compare it to the linear contributions from the relativistic Doppler term, the relativistic potential term, and the lightcone term, following the nomenclature of Fig.\ 1 in \cite{BeutlerDiDio20}. \cite{BeutlerDiDio20} also plots a one-loop contribution to the relativistic potential term, which dominates at $k > 0.1$ $h$ Mpc$^{-1}$. We only plot the linear contribution for simplicity, but, following the relative size of the one-loop term in \cite{BeutlerDiDio20}, we expect that the relativistic potential term will dominate the Doppler bias term at $k \gtrsim 0.1$ $h$ Mpc$^{-1}$. On large scales, however, we find that the Doppler bias can be considerably larger than the relativistic terms, which suggests that proper modelling of the relativistic terms (e.g.,\ to use them to test General Relativity) requires consideration of the Doppler bias.

We performed a number of convergence checks to verify that the calculated power spectra are not affected by numerical issues. 
We found that the results were robust to varying GaPSE settings, aside from the \texttt{cut\_first\_n} and \texttt{cut\_last\_n} arguments. These parameters are necessary when using a window function, because the correlation function at very large scales drops to zero and therefore the FFTLog fails, leading to problematic oscillatory behavior.
We found that cutting the first point and last three points of the correlation function removed the problematic oscillatory behavior for the majority of the power spectrum multipoles, but changing these parameters when making different power spectra is often necessary.

We then compute the signal-to-noise ratio of the difference
between the dipole with $b_D$ and without $b_D$
\begin{equation}
    \left(\frac{S}{N}\right)^2 = \frac{1}{4 \pi^2} V_{\textrm{survey}} \int_{k_{\textrm{min}}}^{k_{\textrm{max}}} dk\, k^2 \frac{[P_1^{b_D}(k) - P_1^{b_D=0}(k)]^2}{\sigma_{P_1}^2(k)}
\end{equation}
where $V_{\textrm{survey}}$ is the volume\footnote{One might wonder why we are choosing $V_{\textrm{survey}}$ to be 14,000 deg$^2$ as opposed to 140 deg$^2$, i.e., the volume probed by the SV1 sample. These biases are, in theory, independent of the volume, and are therefore used with GaPSE, to estimate the two point statistics of the entire sky volume.} probed by a 14,000 deg$^2$ survey at $0.8 < z < 1.0$. We calculate the signal-to-noise separately for the North and South survey regions, since they have quite different values for $b_D$; we assume 5000 deg$^2$ in the North footprint (declination $> 30^{\circ}$) and 9000 deg$^2$ in the South.

For $\sigma_{P_1}(k)$, we use the flat-sky Gaussian covariance estimate from Eq.~7.8 in \cite{BeutlerDiDio20}:
\begin{multline}
    \sigma^2_{P_1^{XY}}(k) = -\frac{9}{10}P_1^{XY}(k)^2 - \frac{18}{35} P_1^{XY}(k) P_3(k) - \frac{23}{70} P_3^{XY}(k)^2 \\ -\frac{20}{77} P_3^{XY}(k) P_5^{XY}(k) - \frac{59}{286} P_5^{XY}(k)^2 \\ + \frac{3}{2 \bar{n}_X} P_0^{YY}(k) + \frac{3}{2 \bar{n}_Y} P_0^{XX}(k)  + \frac{3}{2 \bar{n}_X \bar{n}_Y} \\ + \frac{3}{5 \bar{n}_X} P_2^{YY}(k) + \frac{3}{5 \bar{n}_Y} P_2^{XX}(k)
\end{multline}
This analytic calculation for the dipole covariance neglects window function effects, although we do include their contributions to the multipoles.

\begin{figure}
    \centering
    \includegraphics[width=\textwidth]{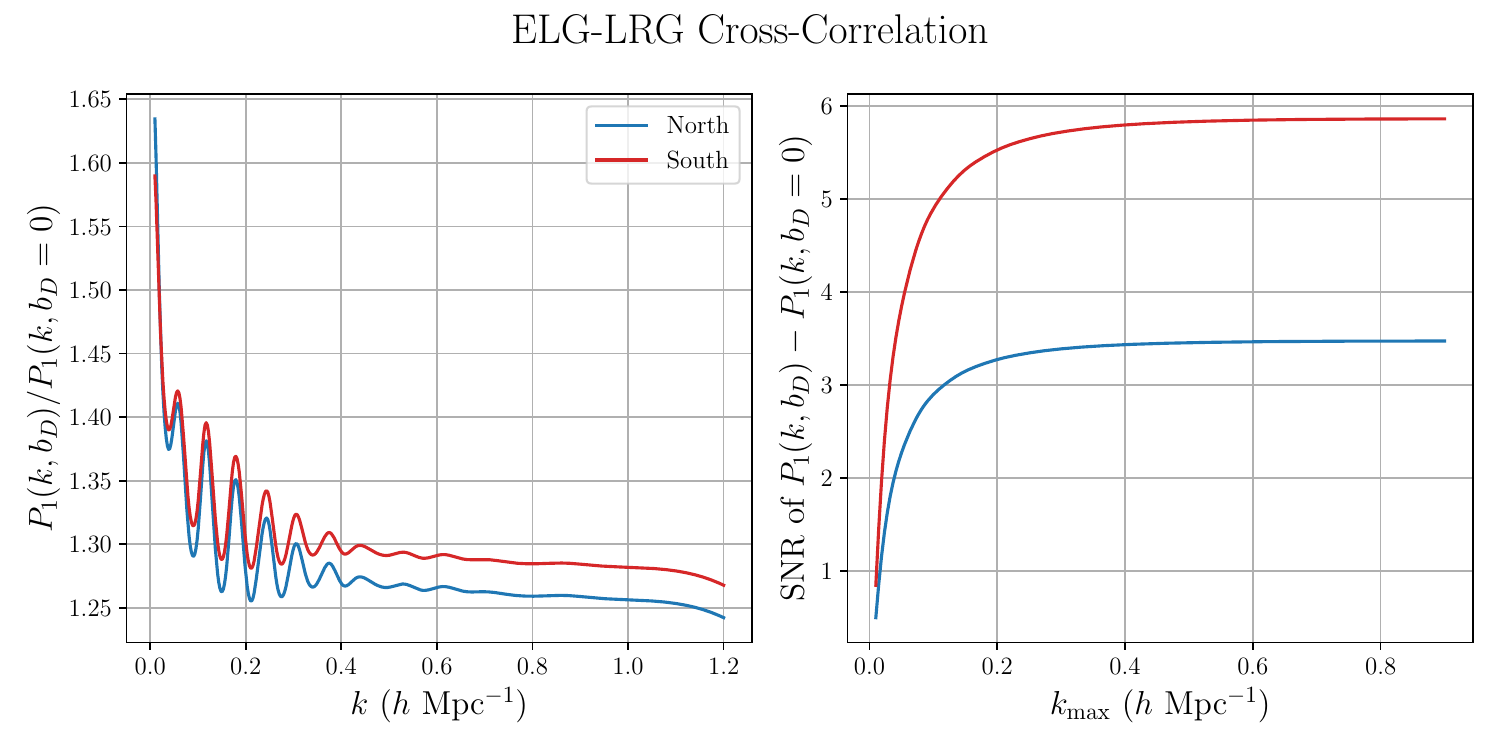}
    \caption{\textit{Left:} fractional impact of the Doppler bias: ratio of the ELG-LRG cross-correlation dipole with $b_D$, to the dipole with $b_D = 0$, in a bin at $0.8 < z < 1$. \textit{Right:} Signal-to-noise ratio of the difference in $P_1(k)$ between the default values of $b_D$ ($-4.7$ in the North and $-5.9$ in the South) and $b_D = 0$, as a function of the smallest scale used $k_{\textrm{max}}$.
    }
    \label{fig:impact_figure}
\end{figure}

We find a cumulative expected signal-to-noise of 7$\sigma$ for the ELG-LRG dipole induced by $b_D$ (Fig.~\ref{fig:impact_figure}), with the South detection (6$\sigma$) considerably larger than the North (3.5$\sigma$). Two differences contribute to the difference in signal strength in the Northern and Southern hemispheres, which both favor the South. First, the South has a larger volume, and second, the South has sharper filter features, which correspond to a stronger Doppler
bias. 

The signal-to-noise is dominated by large, linear scales. It is only reduced 5\% if we use $k_{\textrm{max}} = 0.2$ $h$ Mpc$^{-1}$, and 20\% if we use $k_{\textrm{max}} = 0.1$ $h$ Mpc$^{-1}$.
On the other hand, if we set $k_{\textrm{min}} = 0.02$ $h$ Mpc$^{-1}$, the signal-to-noise is reduced 10\%, and if we set $k_{\textrm{min}} = 0.05$ $h$ Mpc$^{-1}$, the signal-to-noise is reduced 30\%.

This suggests that the Doppler bias could have a measurable impact on the ELG-LRG dipole measured by DESI at $0.8 < z < 1$. We note that there is some systematic uncertainty in our $b_D$ estimates, due to potential incompleteness in the parent ELG sample that we created from DESI SV1. Nevertheless, it is clear that excluding $b_D$ could measurably change theory predictions for $P_1(k)$, potentially
biasing efforts to use the dipole to measure cosmological parameters or test gravity.
At $7\sigma$, the impact of $b_D$ on the DESI ELG-LRG dipole is larger than the 1.6$\sigma$ detection significance for the same cross-correlation forecasted in \cite{BeutlerDiDio20}, which considered the $z \sim 0.3$ DESI BGS sample as a more promising target for the dipole. Indeed, we find a considerably higher total detection significance of the ELG-LRG $P_1(k)$ than \cite{BeutlerDiDio20}, $13\sigma$ in the North and $21\sigma$ in the South.
However, there are several differences between our setup and theirs, which account for this discrepancy.
They assume $s = 0$ and $b_e = 0$, whereas we use the values in Table~\ref{tab:biases_table}. If we set $b_D = 0$ and $b_e = 0$ for both tracers, we get 4.9 and 6.7$\sigma$, and if we also set $s = 0$, we get 
5.7 and 7.6$\sigma$ (the detection significance is slightly higher because $s$ is $<0.4$, i.e.\ it has a negative impact on the overall signal).
Also, the dipole signal in \cite{BeutlerDiDio20} is flat-sky only, i.e.\ they do not count the Newtonian (wide-angle) contribution to the signal as they are only interested in the detectability of the relativistic terms. If we also remove the auto-Newton term from the dipole, we get a detection significance of $2\sigma$ in the North and $3.2\sigma$ in the South.
This is still slightly higher than \cite{BeutlerDiDio20}, but the redshift range considered is different, and, moreover, whether we use the linear power spectrum only (as in this work) or include higher-order terms and an EFT velocity dispersion (as in that work), makes a $\sim30\%$  difference in the signal-to-noise for BGS (their Fig.~10).
Overall, we conclude that our results are broadly consistent with those of \cite{BeutlerDiDio20}, if the same assumptions are made about the galaxy bias parameters. Furthermore, our results suggest that $b_D$ could also have a substantial impact on the DESI BGS sample that they study--although note that we expect $b_D$ to be smaller for BGS since it is not as emission line dominated as the ELGs.

\section{Conclusion and Future Prospects}
\label{conclusion}
By shifting emission lines across sharp features in filter bandpasses, peculiar velocities can have a large impact on the target selection of Emission Line Galaxies in spectroscopic surveys, which uses galaxy colors in broad-band imaging. This leads to an additional contribution to the evolution bias, which we name the Doppler bias, $b_D$. This Doppler bias contributes to relativistic terms in the galaxy power spectrum. We measure the Doppler bias for DESI Emission Line Galaxies (ELGs), using an extended sample of ELGs from DESI's early Survey Validation (SV1) on which we can Doppler shift the galaxies and measure the impact on the main DESI ELG selection. The Doppler bias can be quite large in narrow ranges in redshift, although the overall impact on a broad redshift bin that would be used for galaxy clustering measurements is modest. The Doppler bias depends on the detailed properties of the filters used in the underlying imaging surveys. For DESI, it is thus different between the North and South regions, where the underlying imaging surveys (and thus the filter bandpasses) are different.
We quantify the impact of the Doppler bias on galaxy clustering using the code GaPSE. The largest impact of peculiar velocities is on the imaginary part of the power spectrum for cross-correlations between two tracers of different bias. This translates into the clustering dipole $P_1(k)$, since the Doppler bias breaks inversion symmetry $\vec{r} \rightarrow -\vec{r}$.
DESI LRGs offer an ideal sample for cross-correlation with ELGs in the redshift range of interest, $0.8 < z < 1.0$, with $\Delta b \sim 0.7$. 
We estimate that the Doppler bias contribution changes the ELG-LRG dipole at a $7\sigma$ level ($6\sigma$ in the South and $3.5\sigma$ in the North, where the estimated Doppler bias is smaller due to differences in the filter bandpasses).
While this estimate is affected by systematic uncertainties in the Doppler bias measurement due to potential incompleteness in the SV1 parent sample of ELGs, it suggests that the Doppler bias will have a measurable impact on the multi-tracer dipole and could therefore
affect cosmological measurements or tests of gravity based on the dipole. On the other hand the impact of $b_D$ is much more moderate on the ELG autocorrelation, or the monopole and quadrupole in either the autocorrelation or ELG-LRG cross-correlation.
As rapidly improving datasets will soon allow us to detect the multi-tracer dipole for the first time, our work motivates the inclusion of the Doppler bias in theoretical calculations and modelling that already includes the evolution bias. As we show in Fig.~\ref{fig:two_pt_stats}, the Doppler bias can have a larger impact on the dipole than the relativistic terms at large scales. While we have measured
the Doppler bias for Emission Line Galaxies where we expect the effect to be especially large, it should also exist for other samples of galaxies, such as the DESI Bright Galaxy Sample (BGS) that is considered the most promising to search for relativistic effects in the dipole \cite{BeutlerDiDio20}. Furthermore, rather than attempting to measure $b_D$ from first principles, one could instead use measurements of the dipole to constrain the true value of the Doppler bias, similar to the proposal of \cite{SobralBlanco24} for the evolution bias.

Another interesting future prospect would be exploring the impact of the Doppler bias on parity violation in the galaxy trispectrum or four-point function, as an explanation for the claimed detections of \cite{hou_parity,oliver_parity}.\footnote{But see \cite{PhilcoxEreza,KrolewskiMaySmith} for arguments that the parity-violating signal could be due to residual systematics in the data or simulations used.}
As explored in \cite{paul2024parity},
relativistic terms can contribute to parity violation in the trispectrum, and our Doppler bias would enhance these relativistic terms. 
The magnitude of the parity violation due to the Doppler bias would be highly sample dependent, but is nevertheless crucial to constrain.


Additionally, the dipole measured from the distribution of distant quasars seems to be anomalously large compared to the dipole in the CMB \cite{Secrest_2022,Rameez_2018}. The Doppler bias could contribute an extra term to the classical Ellis-Baldwin formula \cite{EllisBaldwin1984} and thus potentially explain the discrepancy. Further research is necessary to quantify the impact of the Doppler bias in these cases.
Further into the future, Emission Line Galaxies will become increasingly important for large-scale structure measurements, as surveys push to high redshift where galaxies become increasingly faint and strong emission lines offer the best prospects for obtaining redshifts. Euclid will target H$\alpha$ emitters at $z \sim 0.9$--$1.8$ \cite{laureijs2011euclid}; similarly, WFIRST plans to observe both H$\alpha$ emitters and [OIII] galaxies \cite{wfirst}; and, finally, Stage 5 spectroscopic surveys will target Lyman alpha emitters as a key tracer of matter at $z > 2$ \cite{Schlegel22a,Schlegel22b}.
 As a result, understanding the impact of peculiar velocities on galaxy selection will remain important for precision modelling of large-scale galaxy clustering and testing gravity via relativistic effects.

 Interestingly, with the proper engineering of band pass filters, and careful forethought about galaxy targeting, the Doppler bias could realistically be used to probe the radial peculiar velocity field in certain redshifts, though a high degree of precision would be needed in observation, sample completeness, and analysis to usefully probe the large-scale velocity field.

More broadly, biases with similar effects would be useful to study for DESI analysis and future galaxy surveys -- such effects include those of statistical magnitude errors and systematic uncertainties in extinction maps. These would potentially shift galaxies into and out of the selection box, changing the sample, and influencing clustering statistics. Obtaining a relative size of these biases would help present and future surveys determine which effects can most significantly impact their scientific conclusions.
\section{Acknowledgements}
We thank Stephen Bailey, Marco Bonici, Emanuele Castorina, Enea Di Dio, Matt Johnson, Tanveer Karim,   Francisco-Shu Kitaura, Pritha Paul, Will Percival, Hanne Silverans, Licia Verde, Cheng Zhao, and Rongpu Zhou and for useful correspondence, suggestions, and encouragement. We especially thank Dustin Lang for many fruitful discussions throughout the course of this work. 

This work is supported by the Natural Sciences and Engineering Research Council of Canada, the University of Waterloo, and the Perimeter Institute for Theoretical Physics. Research at Perimeter Institute is supported in part by the Government of Canada through the Department of Innovation, Science, and Economic Development of Canada and by the Province of Ontario through the Ministry of Colleges and Universities.

AK was supported as a CITA National Fellow by the Natural Sciences and Engineering Research Council of Canada (NSERC), funding reference \#DIS-2022-568580.

\bibliographystyle{JHEP}
\bibliography{main}

\end{document}